\documentclass[aps,prl,showpacs,groupedaddress,superscriptaddress,twocolumn]{revtex4}

\usepackage{graphicx, amsmath, subfigure}
\usepackage[hypertexnames=false]{hyperref}
\usepackage{color}
\usepackage{float}
\usepackage{chngcntr}

\begin{document}

\preprint{APS/123-QED}


\title{Double-frequency Aharonov-Bohm effect and non-Abelian braiding properties of Jackiw-Rebbi zero-mode}

\author{Yijia Wu}
\affiliation{International Center for Quantum Materials, School of Physics, Peking University, Beijing 100871, China}

\author{Haiwen Liu}
\affiliation{Center for Advanced Quantum Studies, Department of Physics, Beijing Normal University, Beijing 100875, China}

\author{Jie Liu}
\thanks{Corresponding author: jieliuphy@xjtu.edu.cn}
\affiliation{Department of Applied Physics, School of Science, Xian Jiaotong University, Xian 710049, China}

\author{Hua Jiang}
\affiliation{College of Physics, Optoelectronics and Energy, Soochow University, Suzhou 215006, China}

\author{X. C. Xie}
\thanks{Corresponding author: xcxie@pku.edu.cn}
\affiliation{International Center for Quantum Materials, School of Physics, Peking University, Beijing 100871, China}
\affiliation{Beijing Academy of Quantum Information Sciences, Beijing 100193, China}
\affiliation{CAS Center for Excellence in Topological Quantum Computation, University of Chinese Academy of Sciences, Beijing 100190, China}

\date{\today}

\begin{abstract}
Ever since its first proposal in 1976, Jackiw-Rebbi zero-mode has been drawing extensive attention for its charming properties including charge fractionalization, topologically protected zero-energy and possible non-Abelian statistics. We investigate these properties through the Jackiw-Rebbi zero-modes in quantum spin Hall insulator. Though charge fractionalization is not manifested, Jackiw-Rebbi zero-mode's zero-energy nature leads to a double-frequency Aharonov-Bohm effect, implying that it can be viewed as a special case of Majorana zero-mode breaking particle-hole symmetry. Such relation is strengthened since Jackiw-Rebbi zero-modes also exhibit non-Abelian braiding properties in the absence of superconductivity, and the symmetry-protected degeneracy of both Jackiw-Rebbi and Majorana zero-modes is proved to be equally important as the topological gap for their non-Abelian statistics.
\end{abstract}

\maketitle



\textit{Introduction.} Jackiw-Rebbi zero-mode was first raised as the zero-energy soliton solution sandwiched between the positive and negative energy branches of Dirac equation in one spatial dimension \cite{J-R_PRD}. In the presence of such a zero-mode, the total charge of the ``Dirac sea'' is half-integer \cite{J-R_Goldstone_Wilczek_half-charge, J-R_sharp_quantum_observable, J-R_Rajaraman_half-charge_1, J-R_Rajaraman_half-charge_2, J-R_Jackiw_half-charge} due to the charge-conjugation symmetry, which is regarded as another mechanism of charge fractional quantization, in addition to the prestigious fractional quantum Hall (FQH) effect \cite{Laughlin_RMP_Nobel_lecture, J-R_to_FQH-1, Connection_J-R_Majorana-1}. In condensed matter physics, Jackiw-Rebbi zero-mode is closely related to the band topology \cite{Fractionalization_topology, Shun-Qing_Shen_book}. The first famous example is the Su-Schrieffer-Heeger (SSH) model \cite{SSH_model} whose low-energy effective Hamiltonian is equivalent to a 1D topological insulator (TI), and Jackiw-Rebbi zero-mode resides in the domain wall separating topologically distinct phases. Another celebrated example is the Kitaev's chain \cite{Kitaev's_chain} whose effective Hamiltonian is again equivalent to a 1D TI. The difference is that the zero-mode here is self-conjugate due to the superconductivity and therefore a Majorana one. In this vein, Jackiw-Rebbi zero-mode can be regarded as a special case of Majorana zero-mode \cite{Connection_J-R_Majorana-1, Connection_J-R_Majorana-2} in the absence of particle-hole (PH) symmetry.

In the last decade, Jackiw-Rebbi zero-mode was proposed in topological systems including spin ladders \cite{DanielLoss_J-R_non-Abelian-1, DanielLoss_spin_ladder}, Rashba nanowires \cite{DanielLoss_J-R_integer_transmission}, and quantum spin Hall insulator (QSHI) with constriction \cite{DanielLoss_J-R_in_QSH} or external magnetic field \cite{X.L.Qi_proposal_half-charge}. These zero-modes are created or annihilated pairwisely, hence braiding will lead to a non-commutative transformation due to the fermion parity conservation \cite{DanielLoss_J-R_non-Abelian-1, Fermion_parity}. However, on the contrary of its Majorana cousin \cite{IvanovPRL2001}, Jackiw-Rebbi zero-mode's non-Abelian nature 
has not yet been demonstrated in a practical device such as trijunction \cite{MatthewFisher_T-junction, JayDSau_trijunction, Xiong-Jun_Liu_MKP_T-junction, Braiding_error_Y-junction, DanielLoss_spin_ladder} or cross-shaped junction \cite{cross_junction, Chui-Zhen_cross_junction} as has been done for Majorana zero-modes.

Another peculiar property of the Jackiw-Rebbi zero-mode is the one-half charge fractionalization, which has been claimed \cite{X.L.Qi_proposal_half-charge} to be detectable in a pumping process \cite{pump_numeric} or by Coulomb blockade. Recently, a novel 3/2 FQH plateau is observed in single layer 2D electron gas with confined geometry \cite{XiLin_3/2_experiment}. Jackiw-Rebbi zero-mode could be a tentative explanation \cite{XiLin_3/2_experiment} as the confined geometry here may induce inter-edge quasiparticle tunneling, where similar mechanism has been proposed to induce Jackiw-Rebbi zero-mode in QSHI constrictions \cite{DanielLoss_J-R_in_QSH}.

In this Letter, we first show the numerical evidence of Jackiw-Rebbi zero-mode in QSHI heterostructre. Then we investigate the Aharonov-Bohm (AB) effect where a single Jackiw-Rebbi zero-mode is embedded in an AB ring, showing a double-frequency (half flux quantum period) AB oscillation at zero-energy. A comparison with Majorana zero-mode's AB effect supports the aforementioned relation between Jackiw-Rebbi and Majorana zero-mode. We also confirm the non-Abelian braiding properties of Jackiw-Rebbi zero-modes in a cross-shaped QSHI junction. However, such properties are fragile with disorder due to chiral symmetry breaking other than gap closing. Similarly, tiny ``fictitious'' disorder breaking PH symmetry will also destroy Majorana zero-modes' non-Abelian statistics. We draw a conclusion that these zero-modes' degeneracy protected by certain symmetry is indispensable for their non-Abelian properties.


\textit{Lattice model.} The 1D effective Hamiltonian describing four edge channels of QSHI constrictions can be constructed \cite{DanielLoss_J-R_in_QSH} as $ H_{\mathrm{1D}} = v_F \hat{p}_x \rho_z \tau_0 + \Delta_x \rho_x \tau_0 + \Delta_z \rho_z \tau_z + t \rho_x \tau_x $, where the four terms represent the kinetic energy, the spin-orbit interaction (SOI), the Zeeman term, and a spin-conserved inter-edge tunneling term, respectively ($\rho_i$, $\tau_i$ are Pauli matrices working in right-/left-moving spinor and chirality spinor, respectively). The competition between the Zeeman term $\Delta_z$ and the tunneling strength $t$ induces two distinct topological phases and the Jackiw-Rebbi zero-mode resides in the domain wall. The effective Hamiltonian $H_{\mathrm{eff}} = (\Delta_x/t) p_x \pi_x + (\Delta_z-t) \pi_z$ describing topological phase transition has the form of 1D TI ($\pi_i$ are Pauli matrices for real spin). It is worth noting that quantum Hall insulator with two pairs of edge channels \cite{XiLin_3/2_experiment} possesses similar Hamiltonian supporting Jackiw-Rebbi zero-mode.

\begin{figure}[t]
    \centering
    \begin{subfigure}
        
        \hspace{0.03\textwidth} \includegraphics[width=0.42\textwidth]{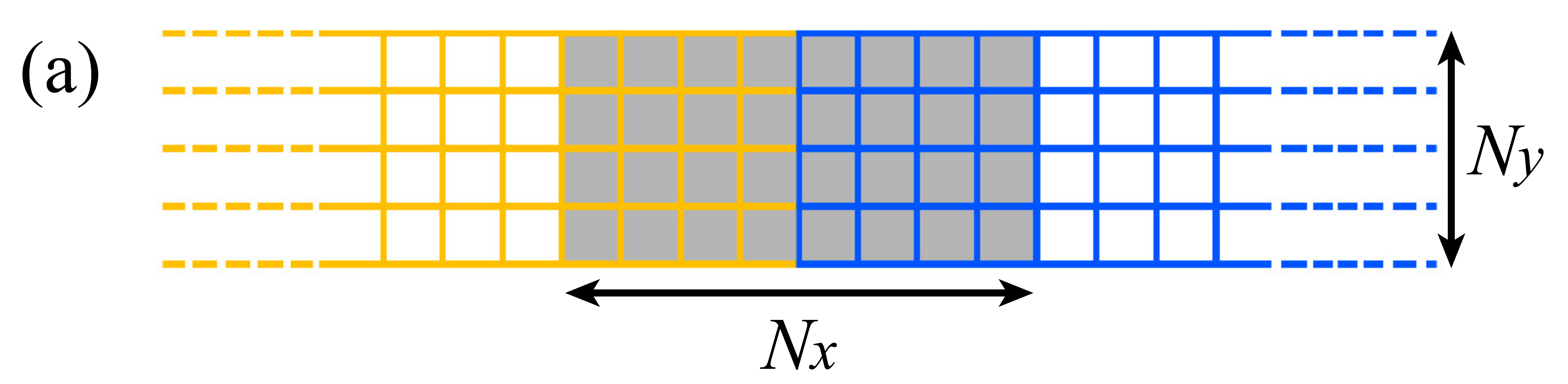}				
	\end{subfigure}
    \begin{subfigure}
        
		\includegraphics[width=0.235\textwidth]{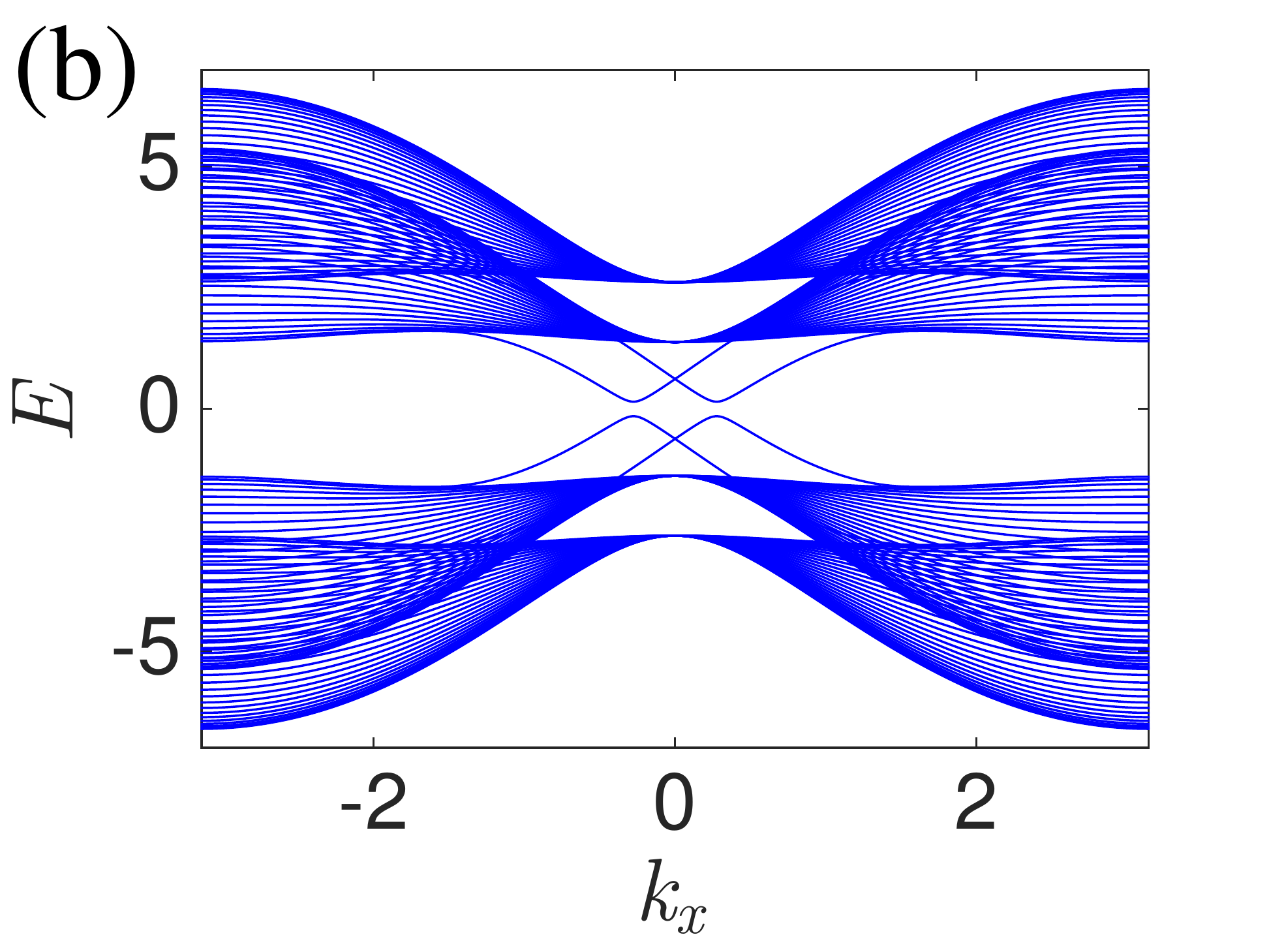}				
	\end{subfigure}
	\begin{subfigure}
	
		\includegraphics[width=0.235\textwidth]{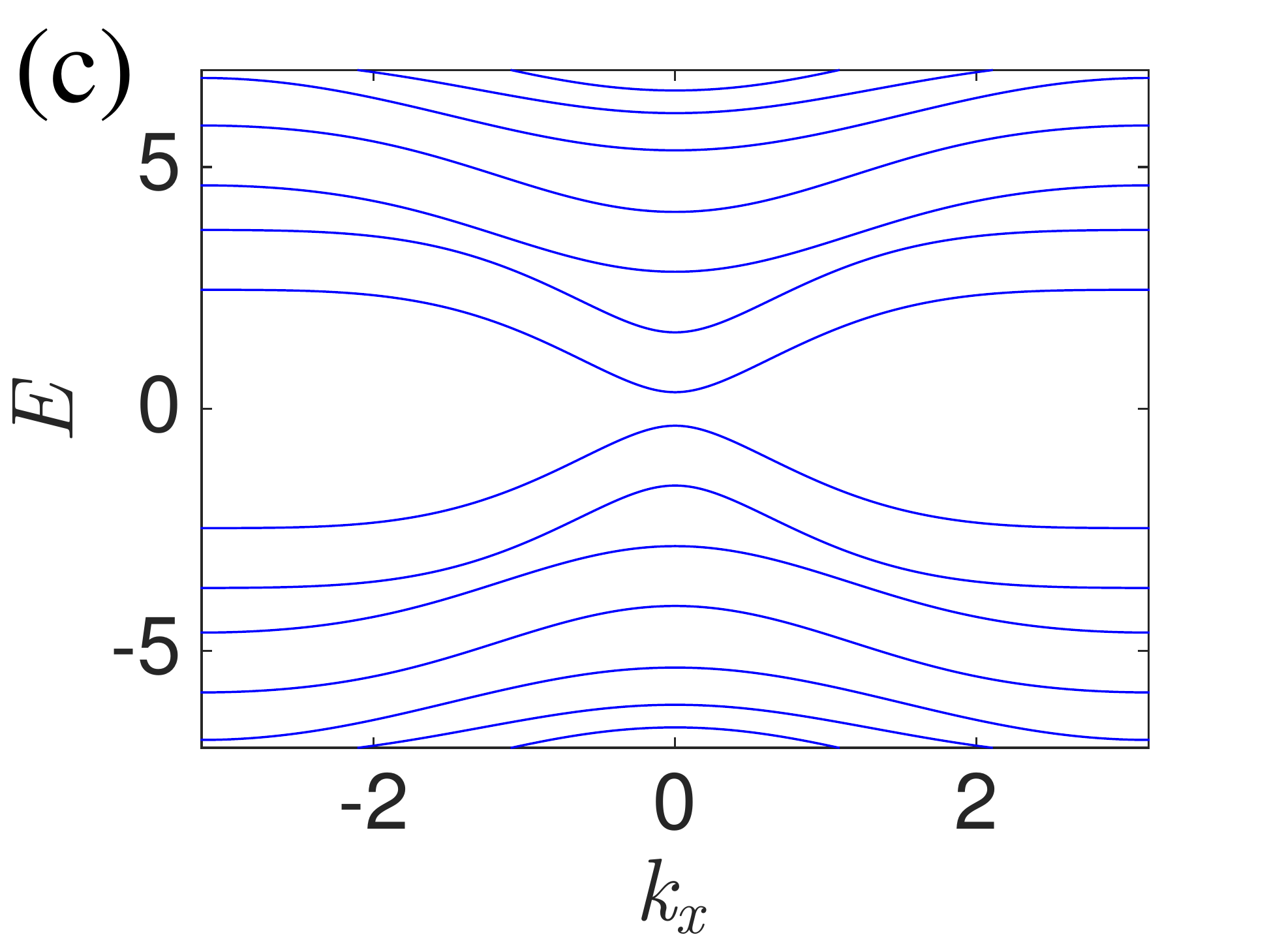}				
	\end{subfigure}
    \begin{subfigure}
        
		\includegraphics[width=0.234\textwidth]{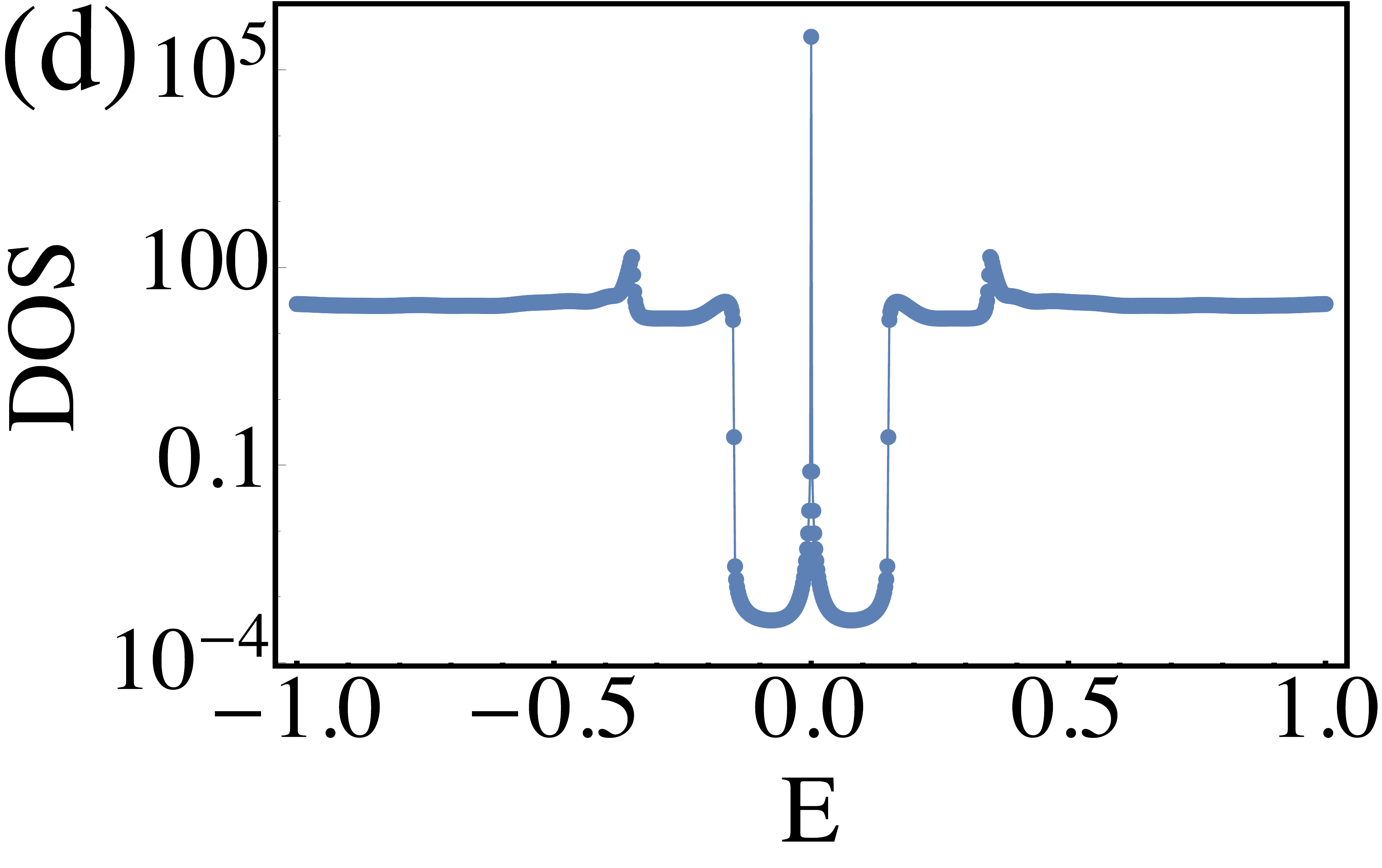}				
	\end{subfigure}
	\begin{subfigure}
    
		\includegraphics[width=0.236\textwidth]{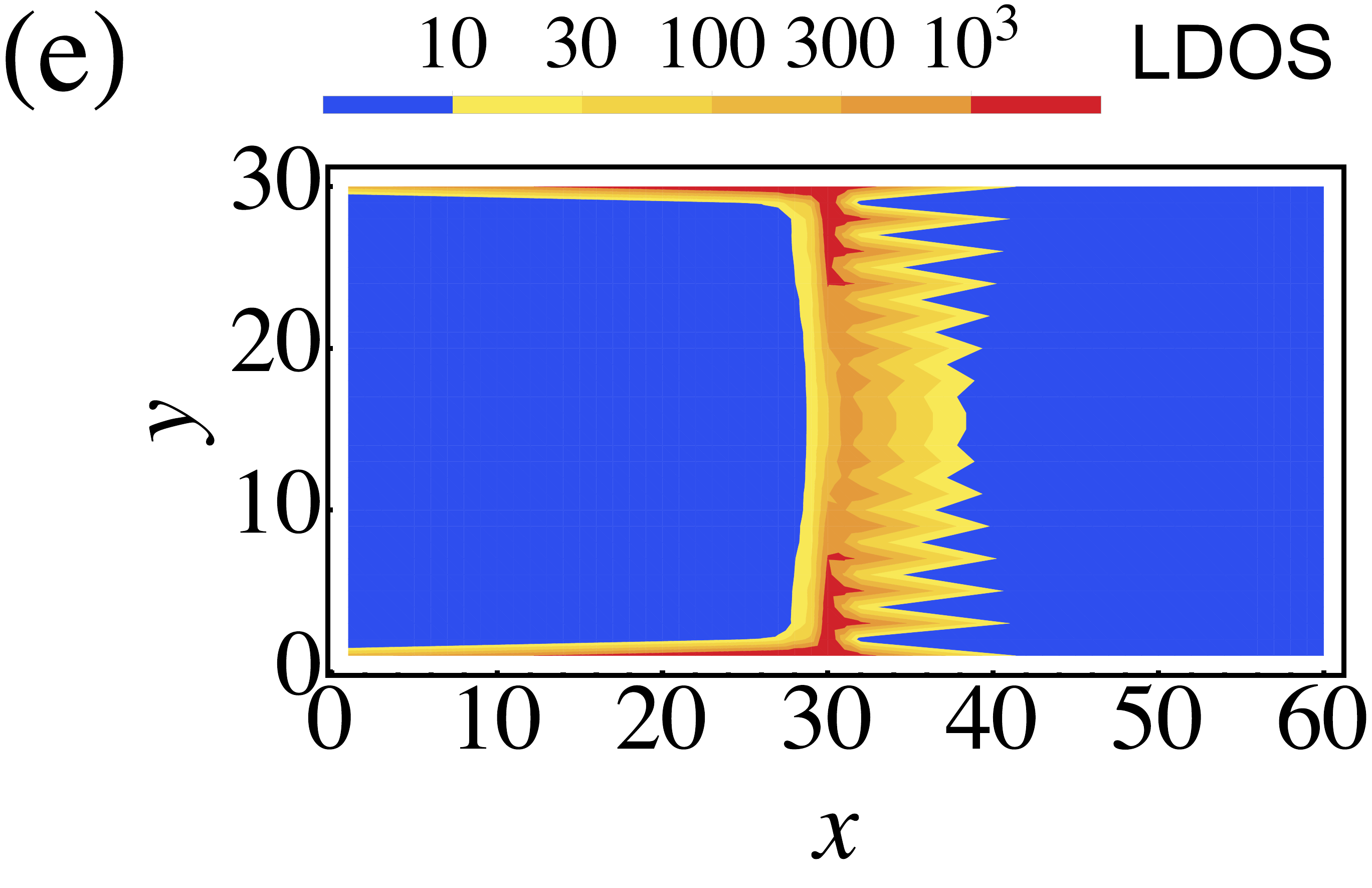}				
	\end{subfigure}
\caption{(a) Schematic diagram of the QSHI heterostructure lattice with finite width $N_y$ [$N_y=30$ through (a)-(e)]. The heterostructure is composed of two semi-infinite QSHIs (yellow and blue, respectively) whose energy spectrums are shown as (b) and (c), respectively. Hamiltonian parameter $A_y=2$ in (b) and there are four edge channels; 
$A_y=40$ in (c) and the edge channels here are destroyed and hence topologically distinct from (b). (d) The DOS inside the shaded region ($N_x=60$) of (a). (e) The local density of states distribution inside the shaded region of (a) at zero-energy $E=0$.}
\label{QSH_heterostructure}
\end{figure}

The 2D Hamiltonian describing QSHI constriction is constructed by adding the Zeeman term $\Delta_z \sigma_0$ and the SOI term $\Delta_x \sigma_0$ into the Bernevig-Hughes-Zhang (BHZ) model \cite{BHZ_model_science, BHZ_model_RMP} ($\sigma_i$ for Pauli matrices):

\begin{equation}
\label{BHZ_model}
H_{\mathrm{QSHI}}(\mathbf{p}) = \begin{pmatrix} 
h(\mathbf{p}) + \Delta_z \sigma_0 & \Delta_x \sigma_0 \\ 
\Delta_x \sigma_0 & h^*(-\mathbf{p}) - \Delta_z \sigma_0  
\end{pmatrix}
\end{equation}

\noindent where $ h(\mathbf{p}) = (A_x p_x \sigma_x - A_y p_y \sigma_y) + (M-B \mathbf{p}^2) \sigma_z $. The discretized version of Eq. (\ref{BHZ_model}) in a square lattice with finite width $N_y$ generally possesses four edge states. However, a strong enough inter-edge tunneling will destroy these edge states and bring about a topologically trivial phase. The tunneling strength is not explicitly shown in Eq. (\ref{BHZ_model}), while it depends on the overlap between the edge states from different edges and therefore can be modulated through the Hamiltonian parameters [e.g., $A_y$, see Fig. \ref{QSH_heterostructure}(b), (c)]. In the method of Green's function, the density of states (DOS) [Fig. \ref{QSH_heterostructure}(d), (e)] of a QSHI heterostructure [Fig. \ref{QSH_heterostructure}(a)] composed of two topologically distinct halves show a subgap zero-energy state localized at the heterostructure's interface. The tunneling strength could also be modulated through $N_y$, which is exactly the earlier proposed QSHI constrictions \cite{DanielLoss_J-R_in_QSH}.

\begin{figure}[t]
    \begin{subfigure}
        
		\includegraphics[width=0.25\textwidth]{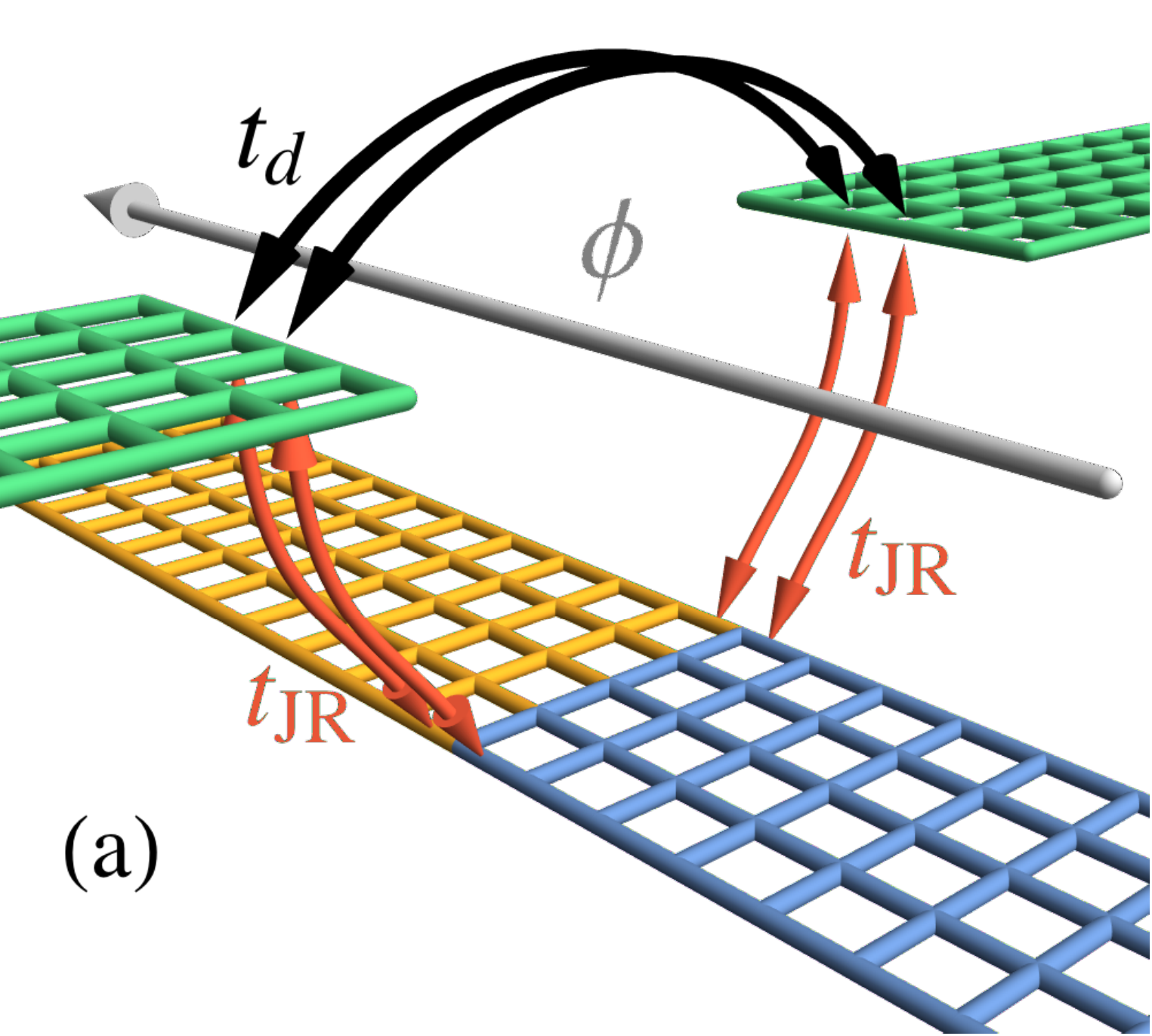}	
	\end{subfigure}
	\begin{subfigure}
	
		\includegraphics[width=0.2\textwidth]{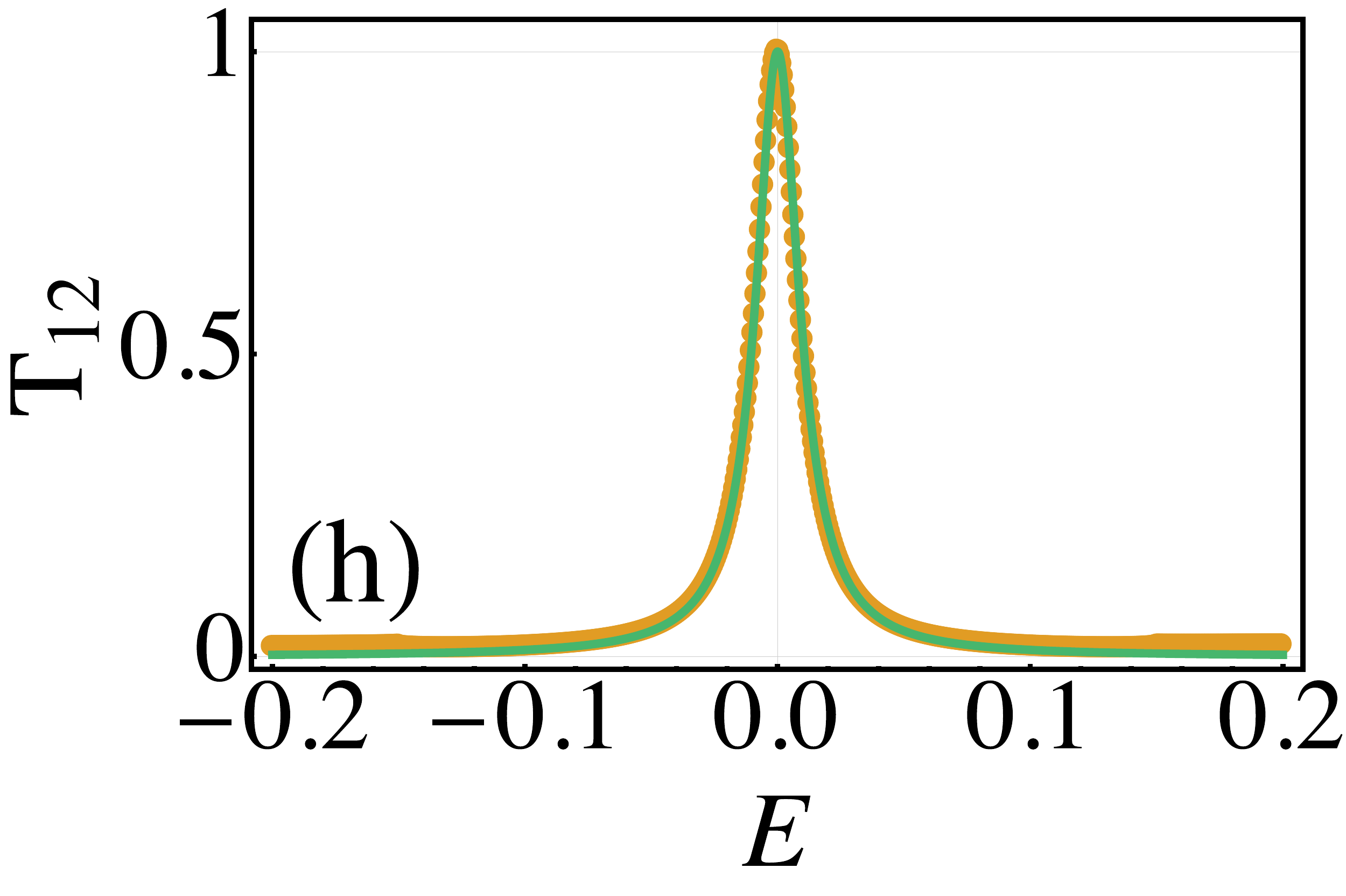}		
	\end{subfigure}
    \begin{subfigure}
        
		\includegraphics[width=0.214\textwidth]{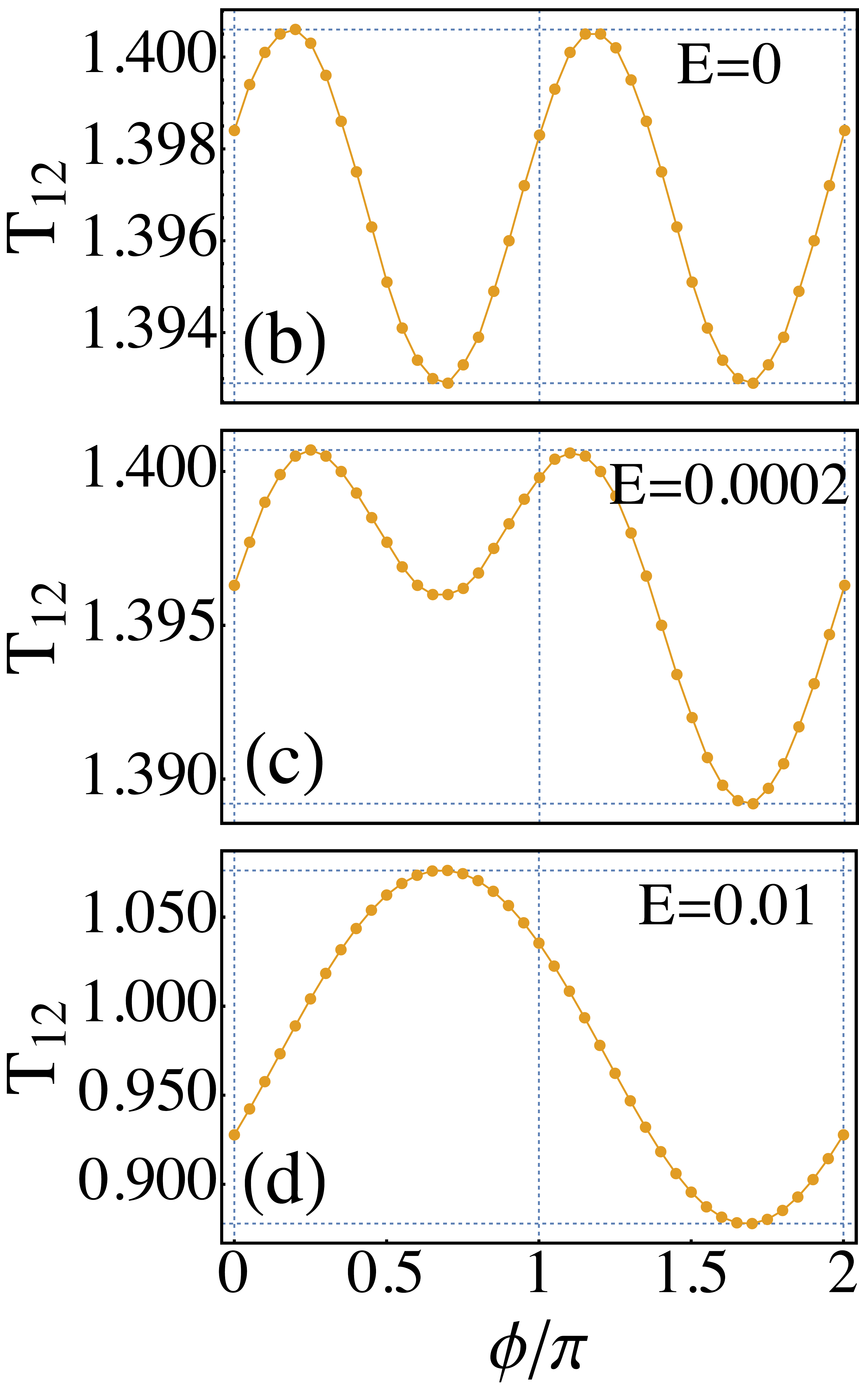}				
	\end{subfigure}
	\begin{subfigure}
    
		\includegraphics[width=0.206\textwidth]{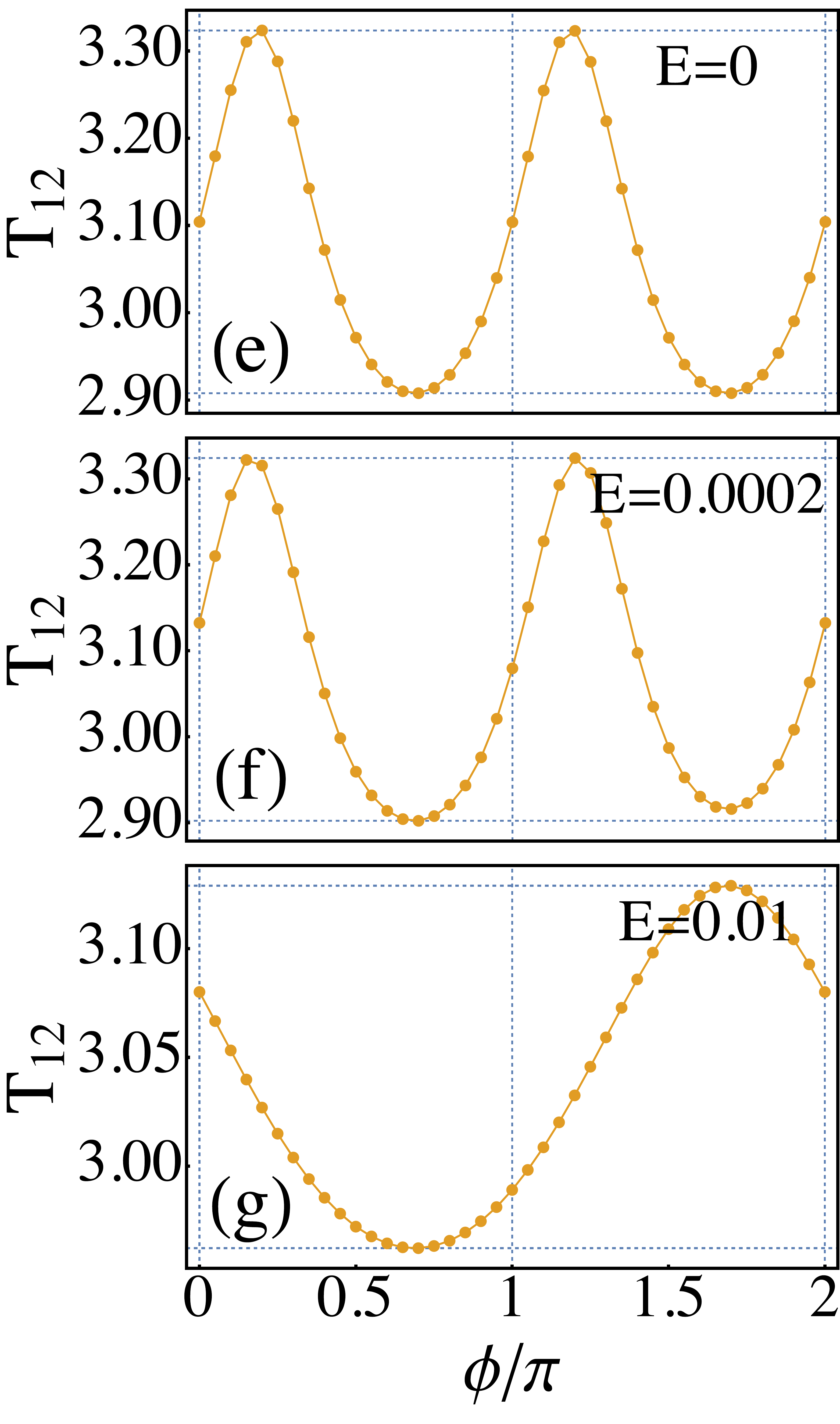}				
	\end{subfigure}  
\caption{(a) Sketch of the AB ring in a lattice model. An AB ring enclosing a magnetic flux $\phi$ is sandwiched between two identical 2D metal leads (shown in green). The upper arm of the ring is the direct hopping between the two leads (black arrows, hopping strength $t_d$), and the lower arm contains a Jackiw-Rebbi zero-mode in a QSHI heterostructure (shown in yellow and blue). The QSHI heterostructure is equally coupled to the two leads (orange arrows, hopping strength $t_{\mathrm{JR}}$). (b)-(d) [(e)-(g)] In the condition of weak (strong) $t_d$, numerical results of $T_{12}$ at fixed energies as (b) [(e)] $E=0$; (c) [(f)] $E=0.0002$; and (d) [(g)] $E=0.01$, respectively. (h) Electron transmission solely induced by the Jackiw-Rebbi zero-mode, where numerical results (orange) can be perfectly fitted by the analytic formula (green) with $\tilde{E} \approx 95E$.}
\label{Numeric_AB_effect}
\end{figure}

\textit{Aharonov-Bohm effect.} Apart from pumping of the domain wall \cite{X.L.Qi_proposal_half-charge, pump_numeric}, transport signature of Jackiw-Rebbi zero-mode is also shown in the electron transmission intermediated by the Jackiw-Rebbi zero-mode \cite{J-R_solitons_RMP, DanielLoss_J-R_integer_transmission}. The pecularities of such an intermediated transmission may be revealed by interfering with a normal electron transmission by direct hopping. Such two-path inteference is exactly the AB effect with a Jackiw-Rebbi zero-mode embedded in a ring geometry [Fig. \ref{Numeric_AB_effect}(a)], where the direct hopping strength between two identical 2D metal leads is $t_d$, and the hopping strength between the leads and the QSHI heterostructure supporting Jackiw-Rebbi zero-mode is $t_{\mathrm{JR}}$. Transmission coefficient between the two leads $T_{12}$ generally depends on the incident electron's energy $E$ as well as the magnetic flux $\phi$ inclosed [$\phi$ is in the unit of $\phi_0/(2\pi)$ below, $\phi_0=hc/e$ is the flux quantum].

The transmission coefficient $T_{12}$ is obtained by numerically calculating the Green's function. For weak $t_d$ [Fig. \ref{Numeric_AB_effect}(b)-(d)], the AB effect shows an unexpected $\pi$-period sinusoidal oscillation in the zero-energy case that the incident electron's energy $E$ matches the energy level of the zero-mode. As the energy $E$ slightly deviates from zero, the oscillation of $T_{12}$ becomes the superposition of a $\pi$-period and a $2\pi$-period sinusoidal functions. The AB effect comes back to the normal $2\pi$-period sinusoidal oscillation at significant non-zero $E$. For strong $t_d$ [Fig. \ref{Numeric_AB_effect}(e)-(g)], though no longer in the simple sinusoidal form, the AB effect still exhibits an unexpected $\pi$-period oscillation for $E=0$, while a $2\pi$-period oscillation for $E \neq 0$.

The scattering matrix ($S$-matirx) theory \cite{KTLaw_Andreev_reflection, Datta_transport} shows that $T_{12}$ has the analytical form as:

\begin{equation}
\label{T_12}
T_{12} = \frac{4\tilde{t}_{d}^{2}\cdot\tilde{E}^{2}+4\tilde{t}_{d}\cos\phi\cdot\tilde{E}+1}{\left[\left(1+\tilde{t}_{d}^{2}\right)\tilde{E}+\tilde{t}_{d}\cos\phi\right]^{2}+1}
\end{equation}

\noindent where $\tilde{t}_d \equiv \frac{t_d}{2v_f}$, and $\tilde{E} \equiv \frac{v_f}{t_0^2} (E-\epsilon_0)$ \cite{SupplementaryMaterial}. In the resonant tunneling condition that $\tilde{E}=0$, $T_{12}$ is reduced to $1/ \left( \tilde{t}_{d}^2 \cos^2\phi +1 \right)$ and therefore exhibits $\pi$-period oscillation. The numerical results could be fitted by Eq. (\ref{T_12}) as $T_{12} = c_0 +c_1 \cdot \frac{4\tilde{t}_{d}^{2}\cdot\tilde{E}^{2}+4\tilde{t}_{d}\cos(\phi+\phi')\cdot\tilde{E}+1}{\left[\left(1+\tilde{t}_{d}^{2}\right)\tilde{E}+\tilde{t}_{d}\cos(\phi+\phi')\right]^{2}+1}$  ($c_0$, $c_1$, and $\phi'$ are constants) \cite{SupplementaryMaterial}. In addition, Eq. (\ref{T_12}) also indicates that $T_{12}=1/(\tilde{E}^2+1)$ in the limit of $\tilde{t}_d \to 0$, which means the transmission solely induced by the Jackiw-Rebbi zero-mode has a peak value of $1$, other than the naively expected value of $1/2$. This analytic result drawn from $S$-matrix is numerically verified [Fig. \ref{Numeric_AB_effect}(h)] and has also been reported in previous research \cite{DanielLoss_J-R_integer_transmission}.


\textit{Comparison with Majorana zero-mode.} If the Jackiw-Rebbi zero-mode embedded in the AB ring is replaced by a Majorana one, then the transmission conductance $G_{12}$ between the two leads is related to the $S$-matrix defined in the Bogoliubov-de Gennes (BdG) basis as $G_{12} = (e^{2}/h) \cdot \left( |S_{12}^{ee}|^{2} - |S_{12}^{he}|^{2} \right)$ ($1, 2$ for lead indices, $e, h$ for electron and hole, respectively, see Supplementary Materials \cite{SupplementaryMaterial}). The explicit form of $G_{12}$ is \cite{transmission_with_Majorana, KTLaw_Andreev_reflection}:

\begin{multline}
\label{T_12_Majorana}
G_{12} = \frac{e^{2}}{h}\cdot\left\{ |S_{12}^{ee}|^{2} - |S_{12}^{he}|^{2}\right\} = \\
\frac{e^{2}}{h} \cdot \frac{-32\tilde{t}_{d} \sin\phi + 8\tilde{t}_{d}\left(1-\tilde{t}_{d}^{2}\right) \cos\phi \cdot \tilde{E} + 4\tilde{t}_{d}^{2}\left(1+\tilde{t}_{d}^{2}\right) \cdot \tilde{E}^{2}}{\left(1+\tilde{t}_{d}^{2}\right)\cdot\left[16+\left(1+\tilde{t}_{d}^{2}\right)^{2}\cdot\tilde{E}^{2}\right]}
\end{multline}

\noindent where $\tilde{t}_d \equiv \frac{t_d}{2v_f}$, $\tilde{E} \equiv \frac{2 v_f}{t_M^2}E$ \cite{SupplementaryMaterial}. In contrast to the Jackiw-Rebbi zero-mode, Eq. (\ref{T_12_Majorana}) indicates that Majorana zero-mode's AB effect oscillates in a $2\pi$-period at both zero-bias ($\tilde{E}=0$) and finite-bias ($\tilde{E} \neq 0$), which is consistent with a former report \cite{Majorana_AB_effect}.

The comparison between the SSH model and the Kitaev's chain hints that Jackiw-Rebbi zero-mode could be viewed as a special case of Majorana zero-mode where the PH symmetry $ H=-\mathcal{P} H^T \mathcal{P}^{-1}$ is absent ($\mathcal{P}$ is PH symmetry operator) \cite{AZclassification_2, Bernevig_Hughes_book}. The electron and hole indices in Eq. (\ref{T_12_Majorana}) are replaced by two electron band indices or sublattice indices (denoted by $\alpha$ and $\beta$) if the Majorana condition is not imposed. Hence the sign differece between electron and hole is absent, and Eq. (\ref{T_12_Majorana}) is modified as $G_{12} = \frac{e^{2}}{h} \{|S_{12}^{\alpha\alpha}|^{2}+|S_{12}^{\beta\alpha}|^{2}\} = \frac{e^{2}}{h}\frac{16}{16+\left(1+\tilde{t}_{d}^{2}\right)^{2} \cdot\tilde{E}^{2}} \cdot \{ \frac{1}{2}+\frac{\tilde{t}_{d}^{2}}{\left(1+\tilde{t}_{d}^{2}\right)^{2}}+\frac{\tilde{t}_{d}^{2}}{4}\tilde{E}^{2} + \frac{\tilde{t}_{d}}{2}\frac{1-\tilde{t}_{d}^{2}}{1+\tilde{t}_{d}^{2}}\cos\phi \cdot \tilde{E} - \frac{\tilde{t}_{d}^{2}}{\left(1+\tilde{t}_{d}^{2}\right)^{2}} \cos2\phi \}$, which qualitatively retrieve the previous consequence that the Jackiw-Rebbi zero-mode exhibits $\pi$-period ($2\pi$-period) AB oscillation at zero-bias (finite-bias).

\begin{figure}[t]
    \centering
    \begin{subfigure}
        
		\includegraphics[width=0.22\textwidth]{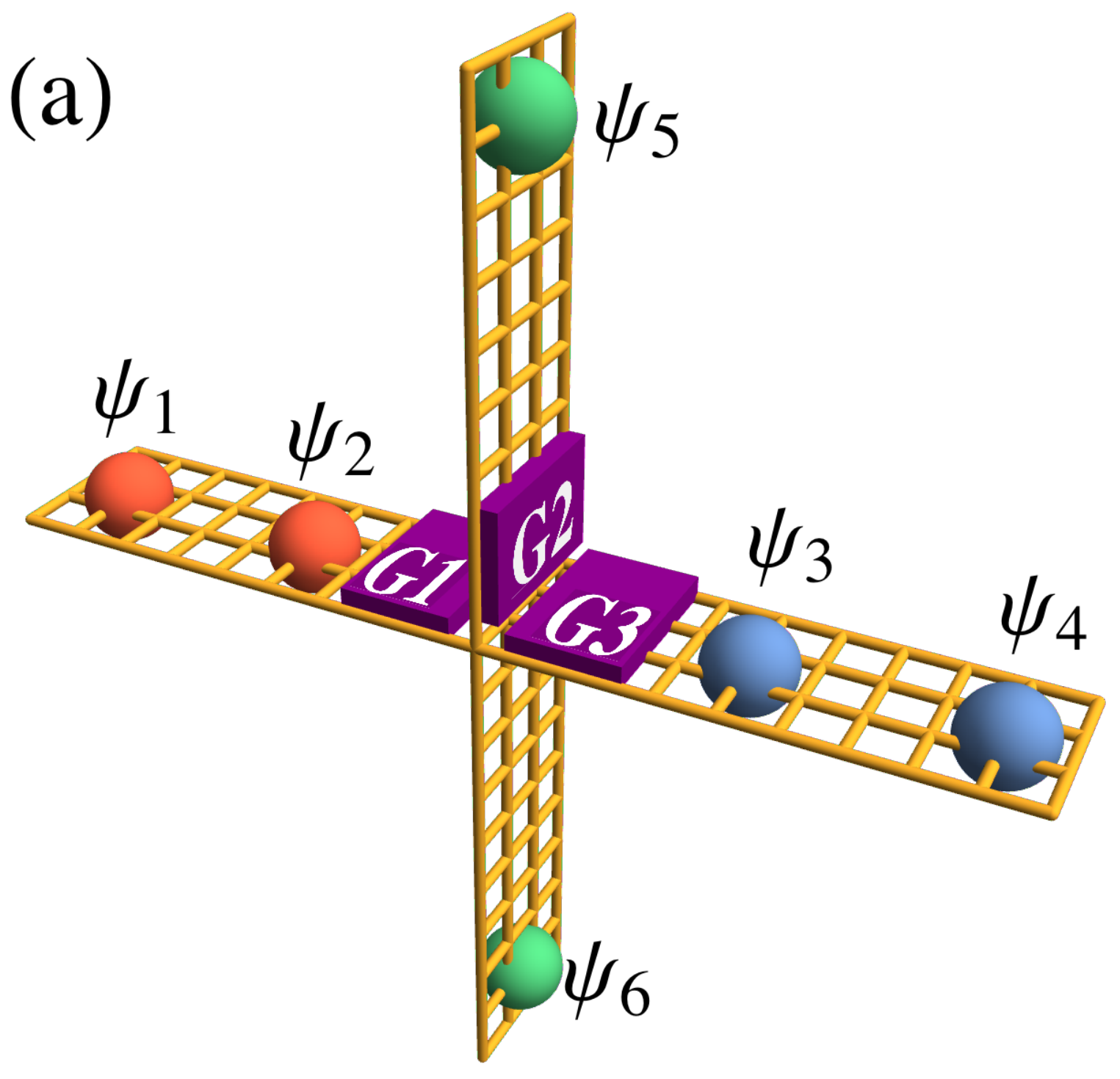}				
	\end{subfigure}
	\begin{subfigure}
	
		\includegraphics[width=0.25\textwidth]{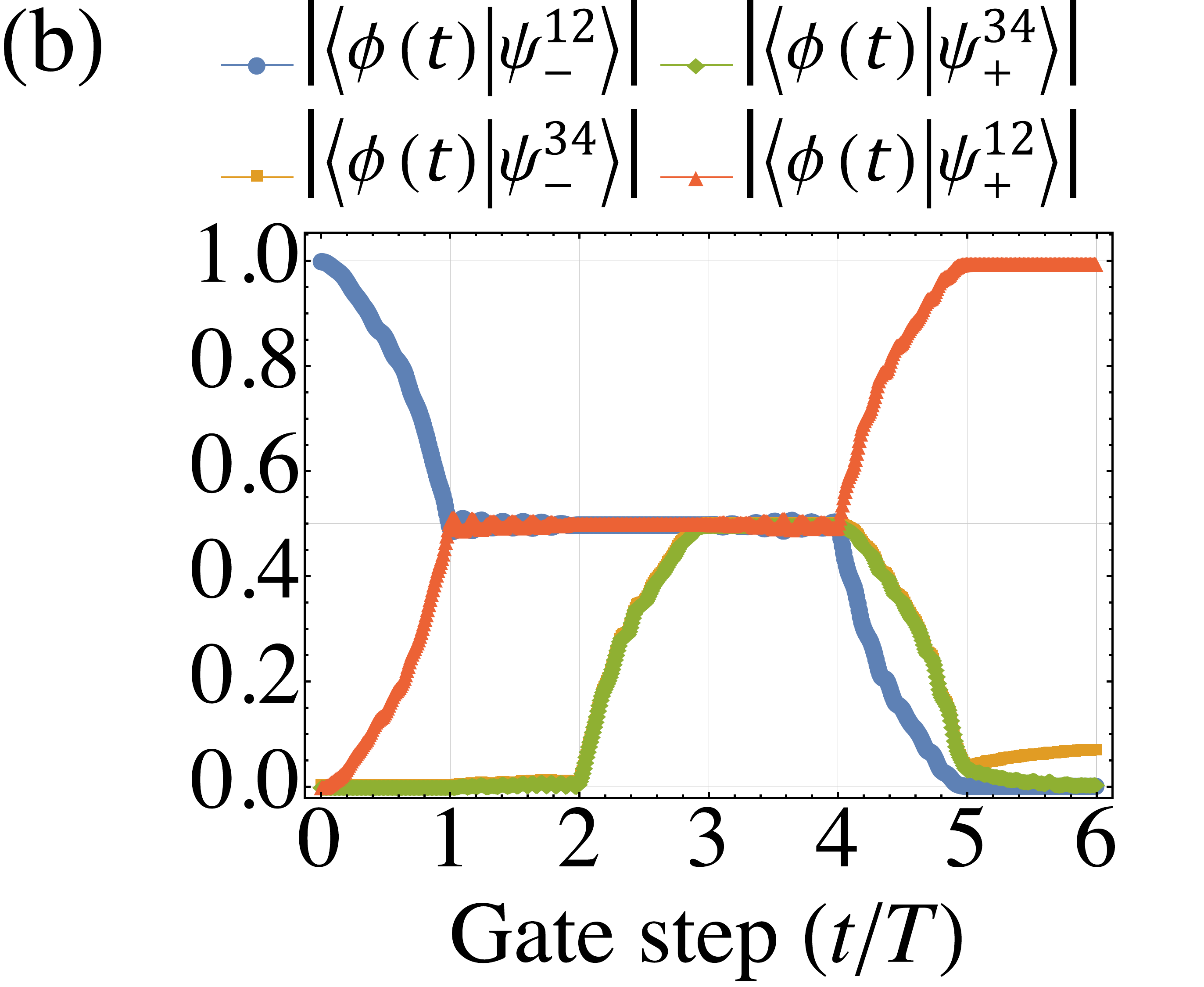}				
	\end{subfigure}
	\begin{subfigure}
	
		\includegraphics[width=0.23\textwidth]{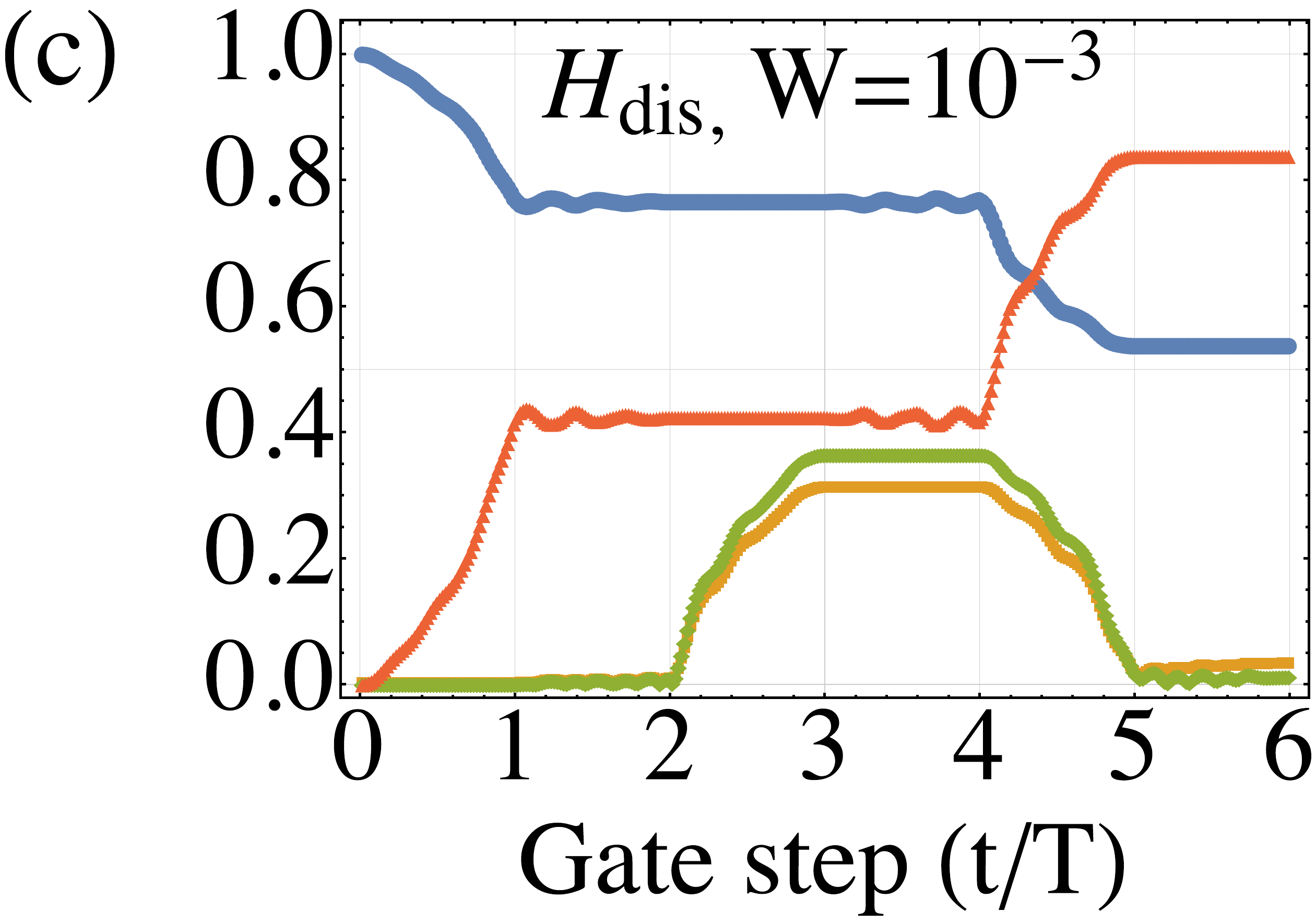}				
	\end{subfigure}
	\begin{subfigure}
	
		\includegraphics[width=0.23\textwidth]{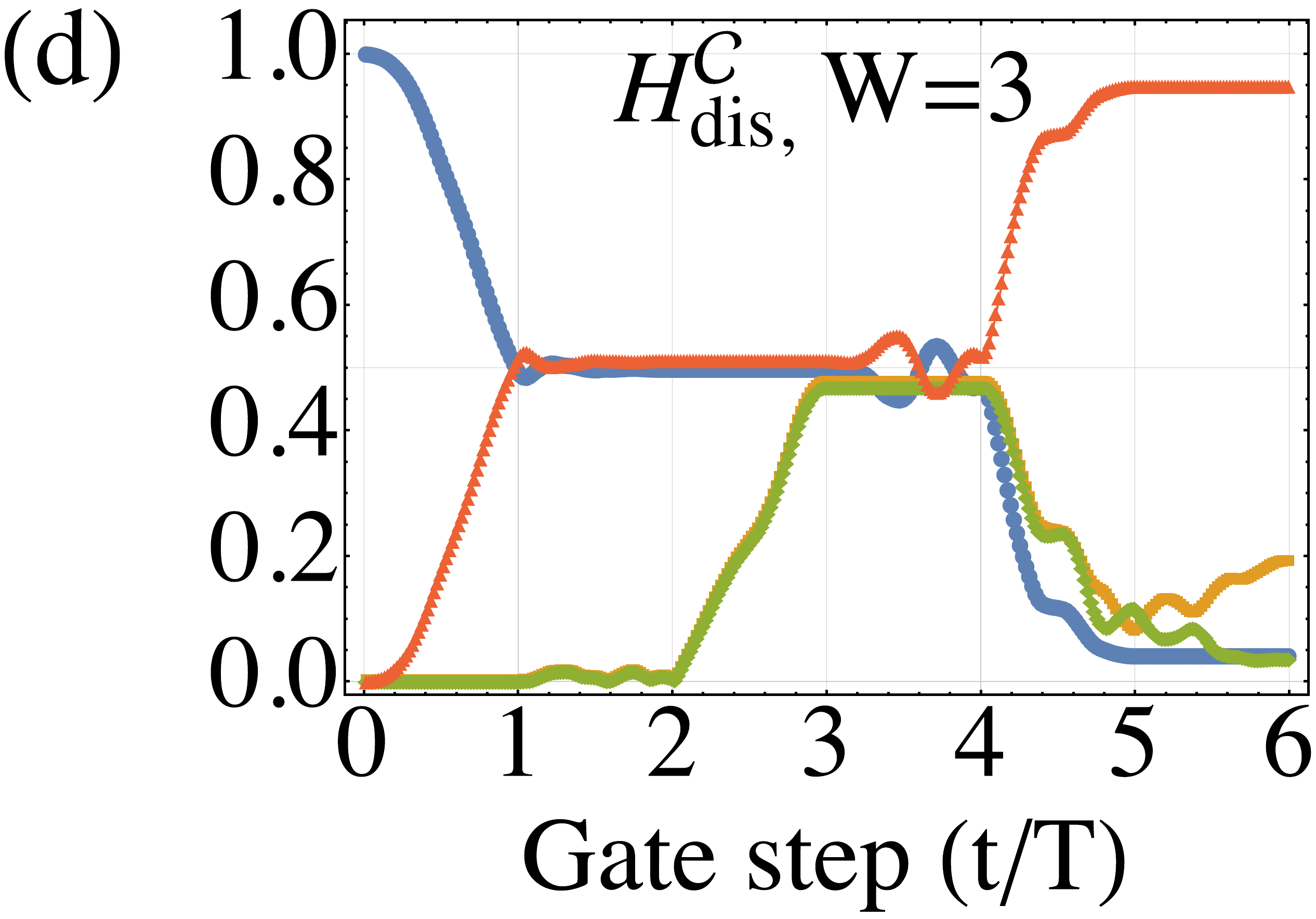}				
	\end{subfigure}
\caption{(a) Sketch of the cross-shaped junction composed of four QSHI arms and three gates. Distribution of the three pairs of Jackiw-Rebbi zero-modes ($\psi_{i=1,2,...,6}$) before braiding are shown. (b)-(d) Evolution of an eigenstate $| \phi(t) \rangle$ as $\psi_2$ and $\psi_3$ are swapped twice (b) in the clean limit; (c) in the presence of chiral symmetry breaking disorder $H_{\mathrm{dis}}$ with $W = 10^{-3}$; (d) in the presence of chiral symmetry conserved disorder $H_{\mathrm{dis}}^{\mathcal{C}}$ with $W=3$. Topological gap $\Delta_b \approx 0.2$, and the coupling energy $\epsilon_{12}, \epsilon_{34} \approx 7\times 10^{-5}$ through (b)-(d).}
\label{cross-junction}
\end{figure}

\textit{Non-Abelian braiding properties.} The similarity between Jackiw-Rebbi and Majorana zero-modes revealed by the AB effect remind us to investigate the possible non-Abelian statistics of Jackiw-Rebbi zero-modes through the cross-shaped junction [Fig. \ref{cross-junction} (a)]. Each of the four arms of the junction is a topologically nontrivial QSHI supporting Jackiw-Rebbi zero-modes, and three gates (G1, G2, and G3) are located near the crossing. If the gate voltage is turned on (off), then the corresponding arm is separated (connected) due to the presence (absence) of the gating potential barrier. Initially, G1 and G3 are turned on while G2 is turned off, hence three pairs of Jackiw-Rebbi zero-modes (denoted by $\psi_{i=1,2,...6}$) are localized at the ends of three divided parts as Fig. \ref{cross-junction} (a). The braiding protocol \cite{cross_junction} takes three steps (time cost for each step is $T$) to swap the spatial positions of $\psi_2$ and $\psi_3$. Firstly, gate voltage G1 is turned off and then G2 is turned on, hence $\psi_2$ is moved to the top of G2. Secondly, G3 is turned off and then G1 is turned on, so now $\psi_3$ is at the left of G1. Thirdly, turning off G2 is followed by turning on G3, as a result $\psi_2$ and $\psi_3$ are swapped.

In the following, we verify that the swapping of $\psi_2$ and $\psi_3$ lead to a geometric phase \cite{vonOppen_braiding} of $\pi$ as $\psi_2 \to \psi_3$ and $\psi_3 \to -\psi_2$, which is exactly the same as Majorana zero-modes \cite{IvanovPRL2001}. In the clean limit, considering the finite-size induced coupling $\epsilon_{2i-1,2i}$ (with phase $\alpha_{2i-1,2i}$) between $\psi_{2i-1}$ and $\psi_{2i}$ ($i=1,2,3$), two Jackiw-Rebbi zero-modes will form symmetric and asymmetric states as $\psi_{\pm}^{12} = \frac{1}{\sqrt{2}} (\psi_1 \pm e^{-i\alpha_{12}} \psi_2)$ \cite{DanielLoss_J-R_non-Abelian-1}. In a whole braiding process swapping $\psi_2$ and $\psi_3$ twice (taking total time of $2 \times 3T$), numerical simulation of the adiabatically approximated time-evolution \cite{SupplementaryMaterial} shows that an eigenstate evolves from $\psi_{-}^{12}$ to $\psi_{+}^{12}$ [Fig. \ref{cross-junction} (b)], which is equivalent to $\psi_2 \to -\psi_2$. Similarly, $\psi_3 \to -\psi_3$ is also confirmed for another eigenstate evolving from $\psi_{-}^{34}$ to $\psi_{+}^{34}$ simultaneously. Consequently, braiding properties $\psi_2 \to \psi_3$ and $\psi_3 \to -\psi_2$ (after swapping $\psi_2$ and $\psi_3$ once) can be drawn from the above results up to a gauge transformation.


\textit{Braiding error with lifting of degeneracy.} Such non-Abelian properties can be destroyed by tiny on-site disorder $ H_{\mathrm{dis}} = \mathrm{diag} \{ V_1(\mathbf{r}), V_2(\mathbf{r}), V_3(\mathbf{r}), V_4(\mathbf{r}) \} $ ($V_i(\mathbf{r})$ uniformly distributed within $[-W/2, W/2]$) where disorder strength $W$ comparable with $\epsilon_{12}, \epsilon_{34}$ but much smaller than the topological gap $\Delta_b$ [Fig. \ref{cross-junction} (c)]. The QSHI constriction [Eq. (\ref{BHZ_model})] only possesses chiral symmetry as $ -H_{\mathrm{QSHI}}(-\mathbf{p}) = \mathcal{C} H_{\mathrm{QSHI}}(\mathbf{p}) \mathcal{C}^{-1}$ with $\mathcal{C} = \pi_y \sigma_y$ ($\pi_y$ is Pauli matrix for real spin) and is classified as AIII symmetry class \cite{AZclassification_1, AZclassification_2}. Strikingly, the non-Abelian properties are preserved until the disorder is strong enough to destruct the topological gap [Fig. \ref{cross-junction} (d)], if the disorder has a chiral symmetry conserved form as $ H_{\mathrm{dis}}^{\mathcal{C}} = \mathrm{diag} \{ V_1(\mathbf{r}), V_2(\mathbf{r}), -V_2(\mathbf{r}), -V_1(\mathbf{r}) \} $ \cite{SupplementaryMaterial}.

Due to the self-conjugation condition $\gamma_i^{\dagger} = \gamma_i$, the Majorana zero-mode's occupation energy term  $\gamma_i^{\dagger} \gamma_i = 1$ is a trivial constant. For Jackiw-Rebbi zero-modes, in contrast, $\psi_i^{\dagger} \neq \psi_i$ due to the lack of PH symmetry and therefore the energy deviation $\Delta_{2i-1,2i}$ in the form of $\psi_i^{\dagger} \psi_i$ (for example, originated from the disorder effect) can be introduced into the effective Hamiltonian as:

\begin{multline}
\label{H_eff_cross_junction}
H_{JR} = \Delta_{12} \psi_{1}^{\dagger}\psi_{1} - \Delta_{12} \psi_{2}^{\dagger}\psi_{2} + \Delta_{34} \psi_{3}^{\dagger}\psi_{3} - \Delta_{34} \psi_{4}^{\dagger}\psi_{4} \\ 
+ \left( \epsilon_{12} e^{i \alpha_{12}} \psi_1^{\dagger} \psi_2 + \epsilon_{34} e^{i \alpha_{34}} \psi_4^{\dagger} \psi_3 + h.c. \right)
\end{multline}

\noindent (widely separated zero-modes $\psi_5$ and $\psi_6$ with negligible coupling are dropped). The four eigenstates of Eq. (\ref{H_eff_cross_junction}) are $\psi_{\pm}^{12} = \frac{1}{\sqrt{2} C_{12}^{\pm}} \{ \psi_1 + e^{-i\alpha_{12}}[ \pm(\widetilde{\Delta}_{12}^2+1)^{1/2} - \widetilde{\Delta}_{12} ] \psi_2 \}$, and $\psi_{\pm}^{34} = \frac{1}{\sqrt{2} C_{34}^{\pm}} \{ \psi_4 + e^{-i\alpha_{34}}[ \pm(\widetilde{\Delta}_{34}^2+1)^{1/2} - \widetilde{\Delta}_{34} ] \psi_3 \}$, respectively ($\widetilde{\Delta}_{2i-1,2i} \equiv \Delta_{2i-1,2i} / \epsilon_{2i-1,2i}$, and $C_{2i-1,2i}^{\pm}$ are normalization constants). In case of non-zero $\Delta_{2i-1,2i}$, numerical simulation confirms that braiding properties $\psi_{2} \to \psi_{3}$ and $\psi_{3} \to -\psi_{2}$ are still valid \cite{SupplementaryMaterial} and thus

\begin{equation}
\label{fidelity}
\left| \langle \phi(t=6T) | \psi_{+}^{12} \rangle \right| = \left( 1+\widetilde{\Delta}_{12}^{2} \right)^{-1/2}
\end{equation}

\noindent where $| \phi(t=0) \rangle = | \psi_{-}^{12} \rangle$. The braiding ``fidelity'' [Eq. ({\ref{fidelity}})] reduces from $1$ to $0$ with the increase of $\widetilde{\Delta}_{12}$ describing the lift of degeneracy. Chiral symmetry breaking disorder $H_{\mathrm{dis}}$ with strength $W \sim \epsilon_{12}$ yileds $\widetilde{\Delta}_{12} \sim 1$ and thus causes significant ``fidelity'' loss. Remarkably, the ``fidelity'' loss could be reduced for stronger coupling $\epsilon_{12}$ as $\widetilde{\Delta}_{12}$ will be relatively small. In contrast, for $H_{\mathrm{dis}}^{\mathcal{C}}$, disorders in opposite signs are imposed on edge states with opposite chirality, hence the energy deviation $\widetilde{\Delta}_{12}=0$ and the non-Abelian properties survive.

Similar ``fidelity'' loss induced by the lift of degeneracy is also confirmed by investigating the disorder effect \cite{SupplementaryMaterial} of Majorana zero-modes in a $p \pm ip$-wave superconducotr (SC), in which the only symmetry is PH symmetry $ H=-\mathcal{P} H^T \mathcal{P}^{-1}$ (D symmetry class \cite{AZclassification_1, AZclassification_2, KitaevChain_symmetry}). The important role of chiral symmetry in Jackiw-Rebbi zero-modes is played by PH symmetry here, which preserves the degeneracy of Majorana zero-modes. In addition, as shown in the Supplementary Materials \cite{SupplementaryMaterial}, the analytical prediction of the ``fidelity'' loss [Eq. (\ref{fidelity})] perfectly fits the numerical simulation results.


\textit{Discussions.} The relation and similarity between Jackiw-Rebbi and Majorana zero-modes are uncovered by AB effect and non-Abelian braiding properties. Though the double-frequency AB oscillation of the Jackiw-Rebbi zero-mode is irrelevant to the charge fractionalization, such effect relies on the resonant condition $\tilde{E}=0$ in which Jackiw-Rebbi zero-mode's zero-energy nature is topologically protected, while such pecularity can be easily removed for an ordinary zero-mode such as a localized state in a quantum dot. As for the non-Abelian properties, realization of the Jackiw-Rebbi zero-mode based braiding still needs a candidate material whose symmetry is robust against disorder. For Majorana zero-mode, the desired degeneracy is more robust since PH symmetry is always presented provided that the superconductivity is not destroyed. Nevertheless, for Majorana-based braiding, considering the adiabatic condition $\epsilon_{12}, \epsilon_{34} \ll 1/T \ll \Delta_b$ \cite{cross_junction} where $\Delta_b$ is the SC gap in the order of $1$meV, it requires larger device scale to reduce the finite-size-induced coupling $\epsilon_{12}, \epsilon_{34}$ and relative low braiding frequency $1/T$. These restrictions could be relaxed for Jackiw-Rebbi zero-modes since superconductivity is no longer required and the bulk gap $\Delta_b$ could be generally larger, which shows the possibility of quantum computation device with higher integration level and higher braiding operation frequency.




\textit{Acknowledgements.} We thank Chui-Zhen Chen, Qing-Feng Sun and Xin Wan for fruitful discussion. This work is financially supported by NBRPC (Grants No. 2015CB921102, No. 2017YFA0303301, and No. 2017YFA0304600) and NSFC (Grants No. 11534001, No. 11504008, No. 11674028, No. 11574245, and No. 11822407).

\bibliography{Jackiw-Rebbi}

\clearpage



\widetext

\begin{center}
\textbf{\large Supplementary Materials for ``Double-frequency Aharonov-Bohm effect and non-Abelian braiding properties of Jackiw-Rebbi zero-mode''}

\vspace{0.15in}

Yijia Wu,$^{1}$ Haiwen Liu,$^{2}$ Jie Liu,$^{3,*}$ Hua Jiang,$^{4}$ and X. C. Xie$^{1,5,6,\dagger}$

\vspace{0.1in}

\textit{$^{1}$International Center for Quantum Materials, School of Physics, Peking University, Beijing 100871, China}

\textit{$^{2}$Center for Advanced Quantum Studies, Department of Physics, Beijing Normal University, Beijing 100875, China}

\textit{$^{3}$Department of Applied Physics, School of Science, Xian Jiaotong University, Xian 710049, China}

\textit{$^{4}$College of Physics, Optoelectronics and Energy, Soochow University, Suzhou 215006, China}

\textit{$^{5}$Beijing Academy of Quantum Information Sciences, Beijing 100193, China}

\textit{$^{6}$CAS Center for Excellence in Topological Quantum Computation, University of Chinese Academy of Sciences, Beijing 100190, China}

\end{center}

\counterwithin{equation}{section} 

\counterwithin{figure}{section} 

\setcounter{equation}{0}
\setcounter{figure}{0}
\setcounter{table}{0}
\setcounter{page}{1}
\makeatletter
\renewcommand{\theequation}{S\arabic{equation}}
\renewcommand{\thefigure}{S\arabic{figure}}
\renewcommand{\bibnumfmt}[1]{[S#1]}
\renewcommand{\citenumfont}[1]{S#1}

\section{Derivation of the $S$-matrix for Jackiw-Rebbi zero-mode's AB effect}

The Hamiltonian describing an AB ring with a Jackiw-Rebbi zero-mode embedded in one arm has the form of:

\begin{equation}
\label{H_AB_effect_J-R}
H_{\mathrm{AB}} = -iv_{f} \sum_{i=1,2} \int_{-\infty}^{+\infty}dx \cdot \psi_{i}^{\dagger}\left(x\right)\partial_{x}\psi_{i}\left(x\right) 
+ \ t_{d} \left[e^{i\phi}\psi_{1}^{\dagger}\left(0\right)\psi_{2}\left(0\right) + h.c. \right]
+ \ t_0 \sum_{i=1,2} \left[\varphi^{\dagger}\left(0\right)\psi_{i}\left(0\right) + h.c. \right] + \epsilon_0 \varphi^{\dagger}\left(0\right)\varphi\left(0\right)
\end{equation}

\noindent where $i=1,2$ is the lead index. The creation operator for the conducting mode (in the metal lead) and the Jackiw-Rebbi zero-mode are denoted as $\psi_{i}^{\dagger}(x)$ and $\varphi^{\dagger}(0)$, respectively. From left to right, the four terms in Eq. (\ref{H_AB_effect_J-R}) are the kinetic energy of the metal leads ($v_f$ the Fermi velocity), direct hopping term (with strength $t_d$) between two metal leads, hopping term (with strength $t_0$) between the Jackiw-Rebbi zero-mode and the metal leads, and the on-site energy (denoted by $\epsilon_{0}$) of the Jackiw-Rebbi zero-mode, respectively. 

The Hamiltonian Eq. (\ref{H_AB_effect_J-R}) is derivated as following. Assuming both these two metal leads in the AB ring contain only one conducting mode per moving direction. Hence the Hamiltonian of the first lead can be written as \cite{KTLaw_Andreev_reflection_supp}:

\begin{equation}
\label{H_L1_original}
H_{L1} = \sum_{\epsilon=L,R} \sum_{\sigma=\uparrow,\downarrow} -iv_{f}\int_{0}^{+\infty}dx\cdot\psi_{1\epsilon\sigma}^{\dagger}\left(x\right)\partial_{x}\psi_{1\epsilon\sigma}\left(x\right)
\end{equation}

\noindent where $v_{f}$ is the Fermi velocity, $\epsilon$ denotes the left-/right-moving mode, and $\sigma$ is the spin index. Assuming $\psi_{1L\sigma}\left(x\right)=\psi_{1R\sigma}\left(-x\right)$ for $x>0$ and supressing the right-moving index $\epsilon = R$, Eq. (\ref{H_L1_original}) is simplified as \cite{KTLaw_Andreev_reflection_supp}:

\begin{equation}
\label{H_L1}
H_{L1} = -iv_{f} \sum_{\sigma=\uparrow,\downarrow} \int_{-\infty}^{+\infty}dx\cdot\psi_{1\sigma}^{\dagger}\left(x\right)\partial_{x}\psi_{1\sigma}\left(x\right)
\end{equation}

\noindent The Hamiltonian for the second lead $H_{L2}$ can be dealed with in the same way. Besides, in the AB ring, the two hopping paths between the tip of these two leads have the form of:

\begin{equation}
\label{H_T_original}
H_{T} = t_{d} \sum_{\sigma=\uparrow,\downarrow} \left[e^{i\phi}\psi_{1\sigma}^{\dagger}\left(0\right)\psi_{2\sigma}\left(0\right) + h.c. \right]
+ \frac{t_0}{\sqrt{2}} \sum_{i=1,2} \sum_{\sigma=\uparrow,\downarrow} \left[\xi_{\sigma}\varphi^{\dagger}\left(0\right)\psi_{i\sigma}\left(0\right) + h.c. \right]
\end{equation}

\noindent Both hopping strength $t_{d}$ and $t_0$ are assumed to be real, $\xi_{\sigma}$ are complex numbers with $|\xi_{\sigma}|=1$ and $\phi$ is the magnetic flux inclosed. Operating a unitary transformation $\begin{cases}
\psi_{i}\left(x\right)=\frac{1}{\sqrt{2}}\left[\xi_{\uparrow}\psi_{i\uparrow}\left(x\right)+\xi_{\downarrow}\psi_{i\downarrow}\left(x\right)\right]\\
\psi_{i}^{'}\left(x\right)=\frac{1}{\sqrt{2}}\left[\xi_{\uparrow}\psi_{i\uparrow}\left(x\right)-\xi_{\downarrow}\psi_{i\downarrow}\left(x\right)\right]
\end{cases}$ (where $i=1,2$) and dropping $\psi_{i}^{'}$ for not participating in the inteference (only contributing a conductance constant), finally we get the full Hamiltonian whose form is exactly Eq. (\ref{H_AB_effect_J-R}) by combining $H_{L1}$, $H_{L2}$, $H_T$ and the on-site energy of the zero-mode.

\begin{figure}[t]
    \centering
    \begin{subfigure}
        
		\includegraphics[width=0.325\textwidth]{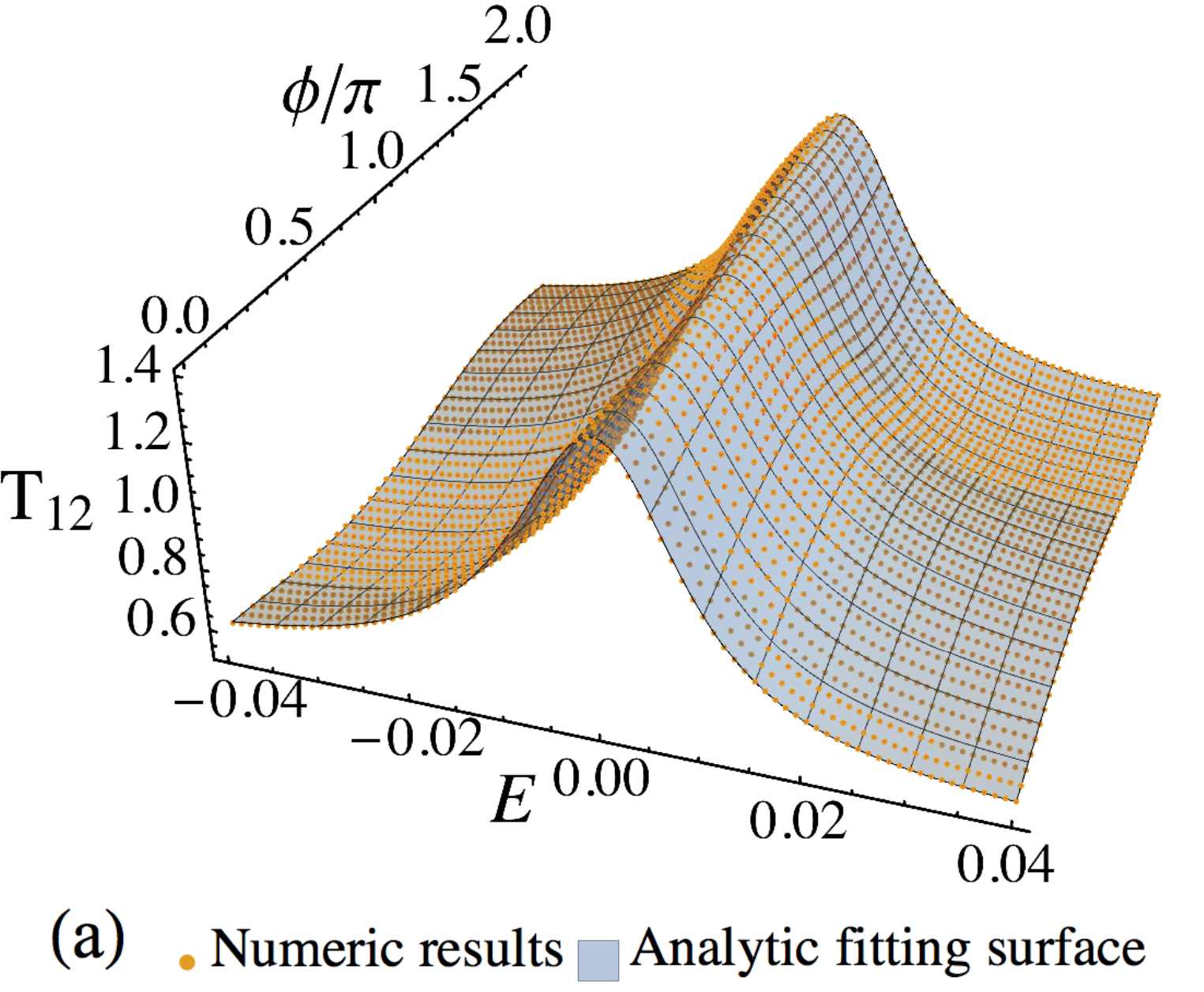}	
	\end{subfigure}
	\begin{subfigure}
	
		\includegraphics[width=0.325\textwidth]{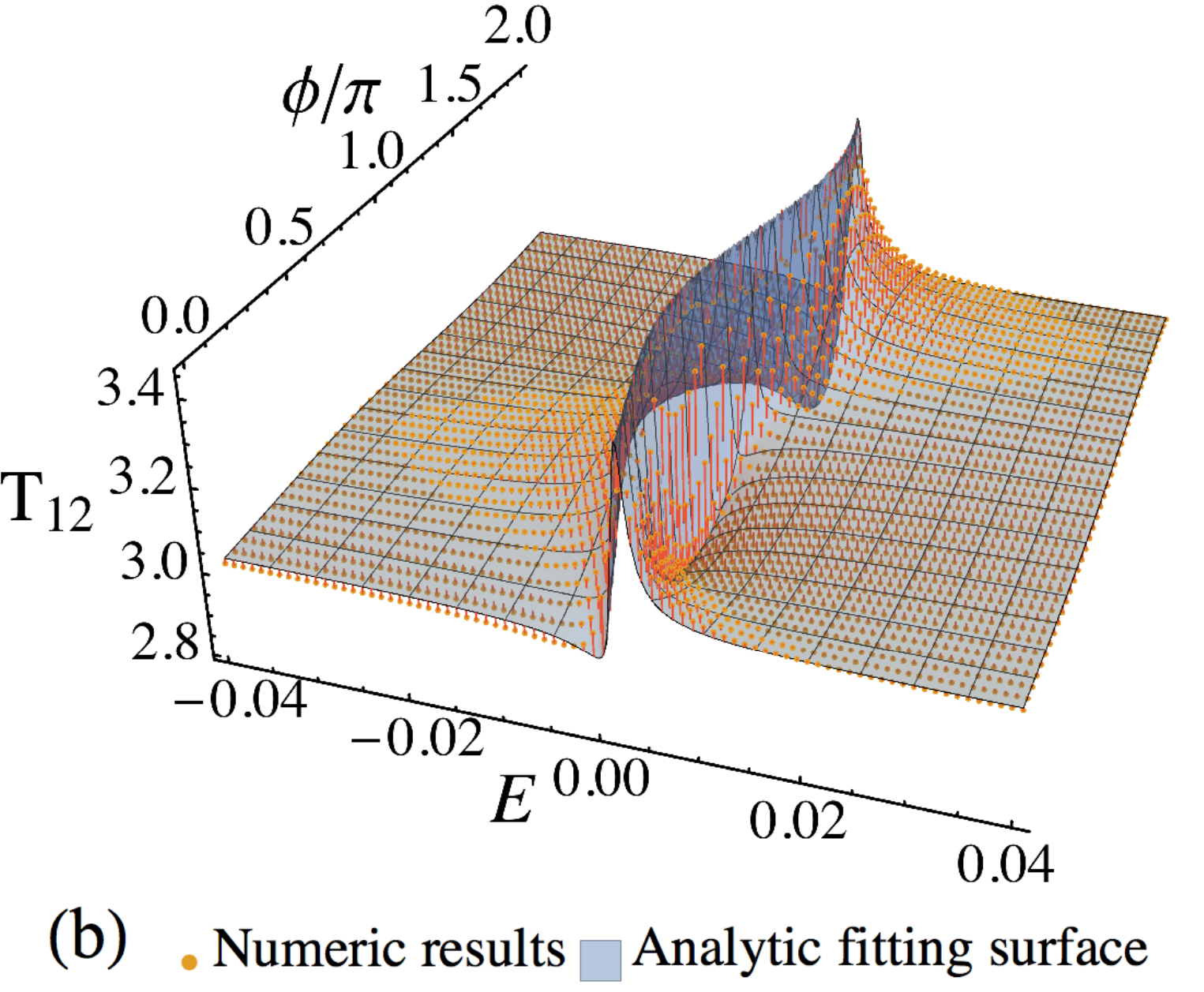}		
	\end{subfigure}
\caption{(a) [(b)] The transmission coefficient $T_{12}$ as a function of the incident electron's energy $E$ and the magnetic flux $\phi$ inclosed, in the condition of weak (strong) $\tilde{t}_d$. Numerical results (obtained by the Green's function) are shown as orange dots, as the fitting formulae (by $S$-matrix) are shown as translucent blue surfaces. Red vertical lines indicate the fitting residues. The fitted parameters in (a) [(b)] are $\tilde{t}_d \approx 0.070$ ($\tilde{t}_d \approx 3.0$) and $\tilde{E} \approx 95E$ ($\tilde{E} \approx 122E$). Fig. \ref{Numeric_AB_effect} (b)-(d) [(e)-(g)] in the main text are replotted from the numerical results shown in (a) [(b)] of this figure.}
\label{AB_effect_fitting}
\end{figure}

Adopting the celebrated Heisenberg's equation of motion (EOM) $i\partial_{t}\hat{O} = [\hat{O},H ]$, we can write down the EOMs for $\psi_1$, $\psi_2$ and $\varphi$ with real space and time variables. Operating the Fourier transform (where $i=1,2$)

\begin{equation}
\begin{cases}
\psi_{iE}\left(x\right)=\frac{1}{2\pi}\int dt\cdot\psi_{i}\left(x,t\right)e^{iEt} \\
\varphi_{E}\left(0\right)=\frac{1}{2\pi}\int dt\cdot\varphi\left(0,t\right)e^{iEt}
\end{cases}
\end{equation}

\noindent and then integrating the EOMs around $x=0$ \{$\varphi_{E}\left(0\right) = [\varphi_{E}\left(0^{+}\right)+\varphi_{E}\left(0^{-}\right) ]/2$, and $\psi_{iE}\left(0\right) = [\psi_{iE}\left(0^{+}\right)+\psi_{iE}\left(0^{-}\right) ]/2$ are inserted\}, finnaly two independent EOMs can be written in a matrix form as

\begin{equation}
\left(\begin{array}{cc}
1-i\tilde{t}_{d}e^{-i\phi} & -1+i\tilde{t}_{d}e^{i\phi}\\
-\frac{i}{2}-\tilde{E} & -\frac{i}{2}-i\tilde{t}_{d}e^{i\phi}\tilde{E}
\end{array}\right)\left(\begin{array}{c}
\psi_{1E}\left(0^{+}\right)\\
\psi_{2E}\left(0^{+}\right)
\end{array}\right) = 
\left(\begin{array}{cc}
1+i\tilde{t}_{d}e^{-i\phi} & -1-i\tilde{t}_{d}e^{i\phi}\\
\frac{i}{2}-\tilde{E} & \frac{i}{2}+i\tilde{t}_{d}e^{i\phi}\tilde{E}
\end{array}\right)\left(\begin{array}{c}
\psi_{1E}\left(0^{-}\right)\\
\psi_{2E}\left(0^{-}\right)
\end{array}\right)
\end{equation}

\noindent where $\tilde{t}_d \equiv \frac{t_d}{2v_f}$, and $\tilde{E} \equiv \frac{v_f}{t_0^2} (E-\epsilon_0)$. The operator at $x=0^{-}$ ($x=0^{+}$) is explained as the incoming (outgoing) mode, since the conducting mode of the lead at $x=0^{-}$ is mapped from the left-moving mode. Therefore, the $S$-matrix defined as $\left(\begin{array}{c}
\psi_{1E}\left(0^{+}\right) \\ \psi_{2E}\left(0^{+}\right) \end{array}\right) = S\left(\begin{array}{c} \psi_{1E}\left(0^{-}\right) \\ \psi_{2E}\left(0^{-}\right) \end{array}\right)$ has the explicit form of (where $\lambda\equiv i\tilde{t}_{d}e^{i\phi}$)

\begin{equation}
S = \frac{1}{\left(1+\tilde{t}_{d}^{2}\right)\tilde{E}+\tilde{t}_{d}\cos\phi+i} \times \left(\begin{array}{cc} \left(1-\tilde{t}_{d}^{2}\right)\tilde{E}-\tilde{t}_{d}\cos\phi & -2\lambda\cdot\tilde{E}-i \\ 2\lambda^{*}\cdot\tilde{E}-i & \left(1-\tilde{t}_{d}^{2}\right)\tilde{E}-\tilde{t}_{d}\cos\phi \end{array}\right) 
\end{equation}

\noindent and $T_{12}$ is the modulus square of the non-diagonal element of the $S$-matrix as $T_{12} = |S_{12}|^2$, which is exactly Eq. (\ref{T_12}) in the main text. As shown in Fig. \ref{AB_effect_fitting}, the numerical results of $T_{12}$ (obtained by the Green's function) could be fitted by the analytic formula as $T_{12} = c_0 +c_1 \cdot \frac{4\tilde{t}_{d}^{2}\cdot\tilde{E}^{2}+4\tilde{t}_{d}\cos(\phi+\phi')\cdot\tilde{E}+1}{\left[\left(1+\tilde{t}_{d}^{2}\right)\tilde{E}+\tilde{t}_{d}\cos(\phi+\phi')\right]^{2}+1}$ [where $c_0$, $c_1$, and $\phi'$ are constants].


\section{Derivation of the $S$-matrix for Majorana zero-mode's AB effect}

The Hamiltonian describing an AB ring with a Majorana zero-mode embedded can be obtained by substituting the last two terms of Eq. (\ref{H_AB_effect_J-R}) by:

\begin{equation}
\label{H_AB_effect_Majorana}
-it_{M}\cdot\eta\left(0\right) \sum_{i=1,2} \left[ \psi_{i}\left(0\right) + h.c. \right] + \epsilon_{M}\cdot\eta^{\dagger}\left(0\right)\eta\left(0\right)
\end{equation}

\noindent where $\eta(0)$ is the Majorana operator. The first term of Eq. (\ref{H_AB_effect_Majorana}) is the coupling (with strength $t_M$) between Majorana zero-mode and the metal leads, as the second term is the on-site energy of the Majorana zero-mode. 

The derivation of the $S$-matrix describing Majorana zero-mode's AB effect is in the same procedure as the Jackiw-Rebbi zero-mode's case. An important difference lies in that the EOMs for $\psi_i$ and $\psi_i^{\dagger}$ are coupled due to the presence of Majorana zero-mode. Therefore the $S$-matrix relates the incoming mode and outgoing mode has the definition of

\begin{equation}
\label{S-matrix_Majorana_definition}
\left(\begin{array}{c}
\psi_{1E}\left(0^{+}\right)\\
\psi_{2E}\left(0^{+}\right)\\
\psi_{1-E}^{\dagger}\left(0^{+}\right)\\
\psi_{2-E}^{\dagger}\left(0^{+}\right)
\end{array}\right)=S\left(\begin{array}{c}
\psi_{1E}\left(0^{-}\right)\\
\psi_{2E}\left(0^{-}\right)\\
\psi_{1-E}^{\dagger}\left(0^{-}\right)\\
\psi_{2-E}^{\dagger}\left(0^{-}\right)
\end{array}\right), \ S\equiv\left(\begin{array}{cccc}
S^{ee} & S^{eh}\\
S^{he} & S^{hh}\\
\end{array}\right)
\end{equation}

\noindent (where $1, 2$ for lead indices, and $e, h$ for electron and hole, respectively) in the Bogoliubov-de Gennes (BdG) basis. The explicit form of the $S$-matrix is shown to be:

\begin{equation}
\begin{footnotesize}
S=C\left(\begin{array}{cccc}
\frac{1-|\lambda|^{4}}{2}\tilde{E}+i\left(1+\lambda\right)\left(1-\lambda^{*}\right) & -i\left(1+\lambda\right)^{2}-\lambda\left(1+|\lambda|^{2}\right)\cdot\tilde{E} & i\left(-1+\lambda^{2}\right) & -i\left(1-\lambda\right)\left(1-\lambda^{*}\right)\\
-i\left(1-\lambda^{*}\right)^{2}+\lambda^{*}\left(1+|\lambda|^{2}\right)\cdot\tilde{E} & \frac{1-|\lambda|^{4}}{2}\tilde{E}+i\left(1+\lambda\right)\left(1-\lambda^{*}\right) & -i\left(1+\lambda\right)\left(1+\lambda^{*}\right) & i\left(-1+\lambda^{*2}\right)\\
i\left(-1+\lambda^{*2}\right) & -i\left(1-\lambda\right)\left(1-\lambda^{*}\right) & \frac{1-|\lambda|^{4}}{2}\tilde{E}+i\left(1-\lambda\right)\left(1+\lambda^{*}\right) & -i\left(1+\lambda^{*}\right)^{2}-\lambda^{*}\left(1+|\lambda|^{2}\right)\cdot\tilde{E}\\
-i\left(1+\lambda\right)\left(1+\lambda^{*}\right) & i\left(-1+\lambda^{2}\right) & -i\left(1-\lambda\right)^{2}+\lambda\left(1+|\lambda|^{2}\right)\cdot\tilde{E} & \frac{1-|\lambda|^{4}}{2}\tilde{E}+i\left(1-\lambda\right)\left(1+\lambda^{*}\right)
\end{array}\right)
\end{footnotesize}
\end{equation}

\noindent The prefactor $C = \frac{2}{\left(1+|\lambda|^{2}\right)\cdot\left[4i+\left(1+|\lambda|^{2}\right)\cdot\tilde{E}\right]}$, besides $\tilde{t}_d \equiv \frac{t_d}{2v_f}$, $\tilde{E} \equiv \frac{2 v_f}{t_M^2}E$, and $\lambda\equiv i\tilde{t}_{d}e^{i\phi}$. With the Majorana zero-mode, the conductance $G_i$ is defined as the derivative of the current inside the $i$th lead \cite{transmission_with_Majorana_supp, Majorana_AB_effect_supp} with respect to the bias $V$:

\begin{equation}
\label{conductance_Majorana}
G_{i} = \frac{dI_{i}}{dV} = \frac{e^{2}}{h}\cdot\left\{ 1-|S_{i1}^{ee}|^{2}-|S_{i2}^{ee}|^{2}+|S_{i1}^{he}|^{2}+|S_{i2}^{he}|^{2}\right\} 
\end{equation}

\noindent The conductance in Eq. (\ref{conductance_Majorana}) can be decomposed into two parts as $G_{i}=G_{ii}-G_{ij}$ ($i\neq j$), where $G_{ii} = \frac{e^{2}}{h} \{ 1-|S_{ii}^{ee}|^{2}+|S_{ii}^{he}|^{2} \}$ is induced by the current which flows out of lead $i$ and then flows back into lead $i$, and $G_{ij} = \frac{e^{2}}{h} \{|S_{ij}^{ee}|^{2}-|S_{ij}^{he}|^{2}\}$ ($i\neq j$) is propotional to the current which flows out of lead $j$ and then flows into lead $i$. Apart from the conductance between two leads $G_{ij}$ [Eq. (\ref{T_12_Majorana}) in the main text], the explicit form of the conductance $G_i$ is

\begin{equation}
G_{1} = \frac{e^{2}}{h} \cdot \frac{16}{\left(1+\tilde{t}_{d}^{2}\right)^{2}\tilde{E}^{2}+16} \cdot \left[ 1 + \frac{2\tilde{t}_{d}}{\left(1+\tilde{t}_{d}^{2}\right)^{2}} \sin\phi\right]
\end{equation}

\noindent and

\begin{equation}
G_{2} = \frac{e^{2}}{h} \cdot \frac{16}{\left(1+\tilde{t}_{d}^{2}\right)^{2}\tilde{E}^{2}+16} \cdot \left[ 1 - \frac{2\tilde{t}_{d}}{\left(1+\tilde{t}_{d}^{2}\right)^{2}} \sin\phi\right]
\end{equation}

It is easy to see that $G_1$ and $G_2$ are in an anticorrelated fashion \cite{Majorana_AB_effect_supp}, and the total conductance $G=G_{1}+G_{2}=\frac{2e^{2}}{h}\cdot\frac{16}{\left(1+\tilde{t}_{d}^{2}\right)^{2}\cdot\tilde{E}^{2}+16}$ is quantized at $\frac{2e^{2}}{h}$ in the zero-energy condition \cite{Majorana_AB_effect_supp}. The total conductance $G$ decays in the manner of $\frac{1}{1+E^{2}}$ for non-zero energy, and the oscillation term $\frac{e^{2}}{h}\cdot\frac{32\tilde{t}_{d} \cdot \sin\phi}{\left(1+\tilde{t}_{d}^{2}\right)\cdot\left[\left(1+\tilde{t}_{d}^{2}\right)^{2}\cdot\tilde{E}^{2}+16\right]}$ is always in the period of $2\pi$.


\section{Numerical simulation for the braiding of Jackiw-Rebbi zero-modes}

The Hamiltonian describing QSHI constriction [Eq. ({\ref{BHZ_model}})] in a square lattice has the form of

\begin{equation}
\label{H_QSHI_realspace}
H_0 = \sum_{i}  \psi_{\mathbf{r}_i}^{\dagger}T_{0}\psi_{\mathbf{r}_i} + \psi_{\mathbf{r}_i}^{\dagger}T_{x}\psi_{\mathbf{r}_i+\delta\mathbf{\hat{x}}} + \psi_{\mathbf{r}_i}^{\dagger}T_{y}\psi_{\mathbf{r}_i+\delta\mathbf{\hat{y}}} + h.c.
\end{equation}

\noindent where $\mathbf{r}_i$ stands for the location of the $i$th lattice site. $T_0$, $T_x$, and $T_y$ are the on-site energy, hopping term along the $x$-direciton, and hopping term along the $y$-direciton, respectively. Each of the four arms in the cross-shaped junction [Fig. \ref{cross-junction} (a) in the main text] can be described by Eq. (\ref{H_QSHI_realspace}), while the hopping term near the crossing controlled by the gate voltages has the form of

\begin{equation}
\label{H_connection_in_cross_junction}
H_{\mathrm{gate}} = \sum_{\langle i, j \rangle} \left( \sum_{\alpha=1,2,3}  g_{\alpha} \psi_{\mathbf{r}_{i,\alpha}}^{\dagger} T_{x} \psi_{\mathbf{r}_{j,c}} + \psi_{\mathbf{r}_{i,4}}^{\dagger} T_{x} \psi_{\mathbf{r}_{j,c}} + h.c. \right)
\end{equation}

\noindent where $\mathbf{r}_{i,\alpha}$ denotes the $i$th lattice site in the $\alpha$th arm ($\alpha=1,2,3,4$), $\mathbf{r}_{j,c}$ denotes the $j$th lattice site at the crossing point, and $\langle i, j \rangle$ means the nearst neighbour. In the numerical simulation, gate voltages G1, G2, G3 are turned on (off) linearly, therefore $g_{\alpha}$ ($\alpha=1,2,3$) in Eq. (\ref{H_connection_in_cross_junction}) is approximated as step functions $g_{\alpha} = 1-n/N$ ($g_{\alpha} = n/N$) with $n=0,1,2,...,N$ ($N$ a large integer).

\begin{figure}[t]
    \centering
    \begin{minipage}{0.45\textwidth} 
        \raggedbottom 
        \includegraphics[width=3in]{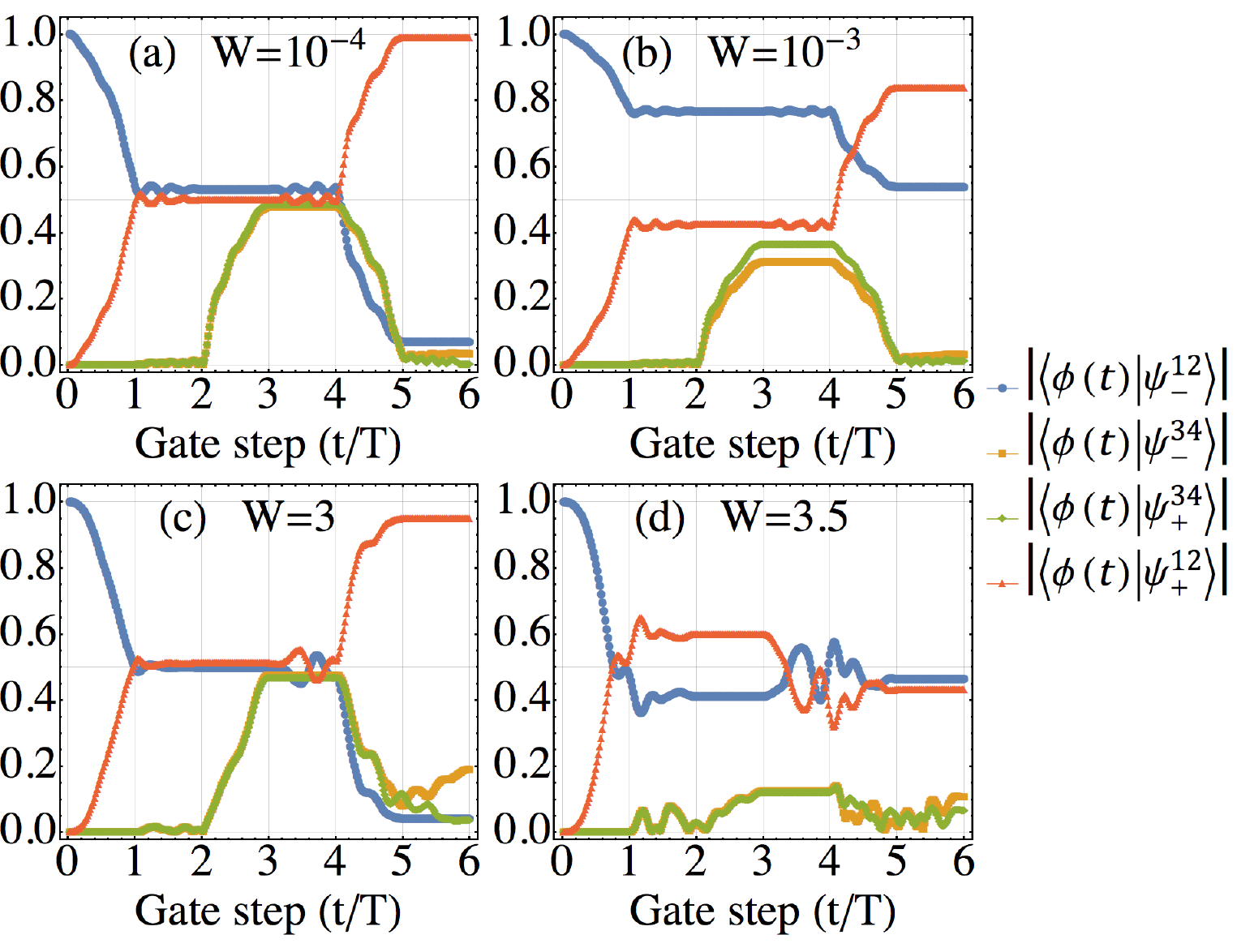}
        
    \end{minipage}%
    \begin{minipage}{0.51\textwidth} 
        \raggedbottom
        \includegraphics[width=3.5in]{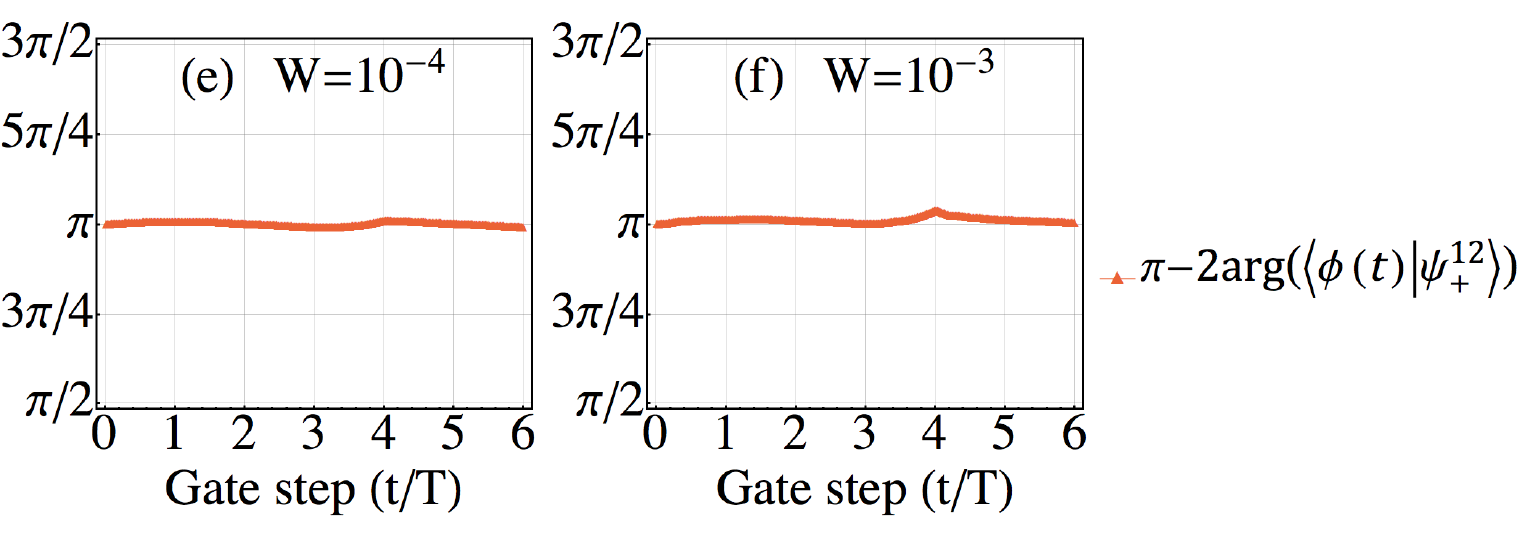}
        
        \vspace{1.17in}	
    \end{minipage}
    				
	\caption{Evolution of the eigenstate $| \phi(t) \rangle$ as two Jackiw-Rebbi zero-modes $\psi_2$ and $\psi_3$ are swapped twice in the presence of disorder. Each gate step takes time of $T=200$, topological gap $\Delta_b \approx 0.2$, and the coupling energy between Jackiw-Rebbi zero-modes $\epsilon_{12}, \epsilon_{34} \approx 7 \times 10^{-5}$. In the presence of chiral symmetry breaking disorder $H_{\mathrm{dis}}$, the modulus and phase angle [Eq. (\ref{phase_angle})] of the projection of $| \phi(t) \rangle$ with disorder strength $W=10^{-4}$ ($W=10^{-3}$) are shown in (a) [(b)] and (e) [(f)], respectively. In the presence of chiral symmetry conserved disorder $H_{\mathrm{dis}}^{\mathcal{C}}$, the modulus of the projection of $| \phi(t) \rangle$ with disorder strength $W=3$ ($W=3.5$) is shown in (c) [(d)]. The non-Abelian braiding properties that $\psi_{-}^{12} \to \psi_{+}^{12}$ is nearly well-preserved for $H_{\mathrm{dis}}$ with $W=10^{-4}$ and $H_{\mathrm{dis}}^{\mathcal{C}}$ with $W=3$, while destructed for $H_{\mathrm{dis}}$ with $W=10^{-3}$ and $H_{\mathrm{dis}}^{\mathcal{C}}$ with $W=3.5$.}
\label{disorder_QSHI}
\end{figure}

The whole braiding Hamiltonian $H_t = H_0 + H_{\mathrm{gate}}$ is time-dependent and the time-evolution operator in the form of $U(t) = \hat{\mathrm{T}} e^{i\int \mathrm{d}t \cdot H(t)}$ ($\hat{\mathrm{T}}$ is the time-ordering operator) is approximated as $U(t) \approx \prod_n e^{i \delta t \cdot H_t}$ due to the step-function approximation. The eigenstate of the junction evolves as $| \phi(t) \rangle = U(t) | \phi(t=0) \rangle$ where $| \phi(t=0) \rangle$ is the initial eigenstate (before braiding). As the braiding protocol stated in the main text, each braiding step takes time of $T$. The adiabatic condition is satisfied when the excitation energy $\sim 1/T$ will not give rise to energy level transition. There are two energy scales in the QSHI cross-shaped junction, the topological gap $\Delta_b$, and the coupling between Jackiw-Rebbi zero-modes $\epsilon_{12}, \epsilon_{34}$. In both Fig. \ref{cross-junction} (b)-(d) in the main text and Fig. \ref{disorder_QSHI}, $\Delta_b \approx 0.2$ and $\epsilon_{12}, \epsilon_{34} \approx 7\times 10^{-5}$, so we choose $\delta_t=0.1$ and $N=1000$, hence the time cost in each braiding step $T = 2\times N \delta_t = 200$ satisfies the adiabatic condition as $\Delta_b \gg 1/T \gg \epsilon_{12}$ \cite{cross-junction_supp}.

In the presence of chiral symmetry breaking disorder  $ H_{\mathrm{dis}} = \mathrm{diag} \{ V_1(\mathbf{r}), V_2(\mathbf{r}), V_3(\mathbf{r}), V_4(\mathbf{r}) \} $, as shown in Fig. \ref{disorder_QSHI} (a), (b), the non-Abelian properties of Jackiw-Rebbi zero-modes are destructed for disorder strength $W$ comparable with $\epsilon_{12}, \epsilon_{34}$ but much smaller than the topological gap $\Delta_b$. Nevertheless, $|\langle \phi(t=6T) | \psi_{-}^{12} \rangle|^2 + |\langle \phi(t=6T) | \psi_{+}^{12} \rangle|^2 = 1$ is still satisfied [Fig. \ref{disorder_QSHI} (a), (b), where the eigenstate before braiding $| \phi(t=0) \rangle = | \psi_{-}^{12} \rangle$], indicating the eigenstate after braiding $| \phi(t=6T) \rangle$ still lives in the Hilbert space spanned by $| \psi_1 \rangle$ and $| \psi_2 \rangle$ (other than mix with other states such as $| \psi_3 \rangle$). Therefore, it is quite reasonable to assume that the whole braiding process swapping $\psi_2$ and $\psi_3$ twice in the presence of disorder give rise to $\psi_2 \to e^{i\theta} \psi_2$, in other words, $| \phi(t=6T) \rangle = \frac{1}{\sqrt{2} C_{12}^{-}} \{ \psi_1 + e^{-i\alpha_{12}}[ -(\widetilde{\Delta}_{12}^2+1)^{1/2} - \widetilde{\Delta}_{12} ] e^{i\theta} \psi_2 \}$. As a result,

\begin{equation}
\label{after_braiding}
\langle \phi(t=6T) | \psi_{+}^{12} \rangle = \frac{1-e^{-i\theta}}{2\sqrt{\widetilde{\Delta}_{12}^{2}+1}} = \frac{\sin{(\theta/2})}{\sqrt{\widetilde{\Delta}_{12}^{2}+1}} \cdot e^{i\frac{\pi-\theta}{2}}
\end{equation}

\noindent and therefore

\begin{equation}
\label{phase_angle}
\theta = \pi - 2\arg \langle \phi(t=6T) | \psi_{+}^{12} \rangle
\end{equation}

\noindent $\theta=\pi$ is verified by the numerical results [Fig. \ref{disorder_QSHI} (e), (f)]. Consequetly, the braiding property $\psi_2 \to -\psi_2$ after swapping $\psi_2$ and $\psi_3$ twice is still valid even in the presence of disorder. Similarly, the braiding property $\psi_3 \to -\psi_3$ can be demonstrated in the same way. These braiding properties indicate that swapping $\psi_2$ and $\psi_3$ once lead to the general form of $\psi_2 \to e^{i\theta_1}\psi_3$ and $\psi_3 \to e^{i\theta_2}\psi_2$ where $e^{i\theta_1}e^{i\theta_2} = -1$. Adopting a gauge transformation imposing $e^{i\theta_1}=1$, then the braiding properties $\psi_2 \to \psi_3$ and $\psi_3 \to -\psi_2$ is exactly the same as Majorana zero-modes \cite{IvanovPRL2001_supp}.

On the contrary, as shown in Fig. \ref{disorder_QSHI} (c), (d), if the on-site disorder has the chiral symmetry conserved form as $ H_{\mathrm{dis}}^{\mathcal{C}} = \mathrm{diag} \{ V_1(\mathbf{r}), V_2(\mathbf{r}), -V_2(\mathbf{r}), -V_1(\mathbf{r}) \} $ (satisfying $ -H_{\mathrm{dis}}^{\mathcal{C}} = \mathcal{C} H_{\mathrm{dis}}^{\mathcal{C}} \mathcal{C}^{-1}$), then the non-Abelian properties are preserved until the disorder is strong enough to close the topological gap. $|\langle \phi(t=6T) | \psi_{-}^{12} \rangle|^2 + |\langle \phi(t=6T) | \psi_{+}^{12} \rangle|^2$ is significantly smaller than $1$ in Fig. \ref{disorder_QSHI} (d), indicating the gap is destructed and the zero-modes are mixed with the bulk states.


\section{Braiding properties of Majorana zero-modes in the presence of disorder}

Majorana zero-modes' braiding is performed with the same protocol and in the same shape of junction as Jackiw-Rebbi zero-modes, where the difference is that the cross-shaped junction here is composed of $p \pm ip$-wave SC supporting Majorana zero-modes. The Hamiltonian of $p \pm ip$-wave SC in the BdG basis

\begin{equation}
\label{p+ip_SC}
H_{p \pm ip\mathrm{SC}}(\mathbf{p}) = \frac{1}{2} \times \left( c_{\mathbf{p}}^{\dagger}, c_{\mathbf{-p}} \right) 
\left[ \Delta (p_x \sigma_x + p_y \sigma_y) + (\frac{\mathbf{p}^2}{2m}-\mu) \sigma_z \right]
\left(\begin{array}{c}
c_{\mathbf{p}}\\
c_{\mathbf{-p}}^{\dagger}
\end{array}\right)
\end{equation}

\noindent possesses PH symmetry as $ -H(-\mathbf{p}) = \mathcal{P} H^T(\mathbf{p}) \mathcal{P}^{-1}$ with $\mathcal{P}=\sigma_x$, and hence is in the D symmetry class (both TR and chiral symmetries are absent, same symmetry class as Kitaev's chain with complex SC pairing).

\begin{figure}[t]
    \centering
	\includegraphics[width=0.45\textwidth]{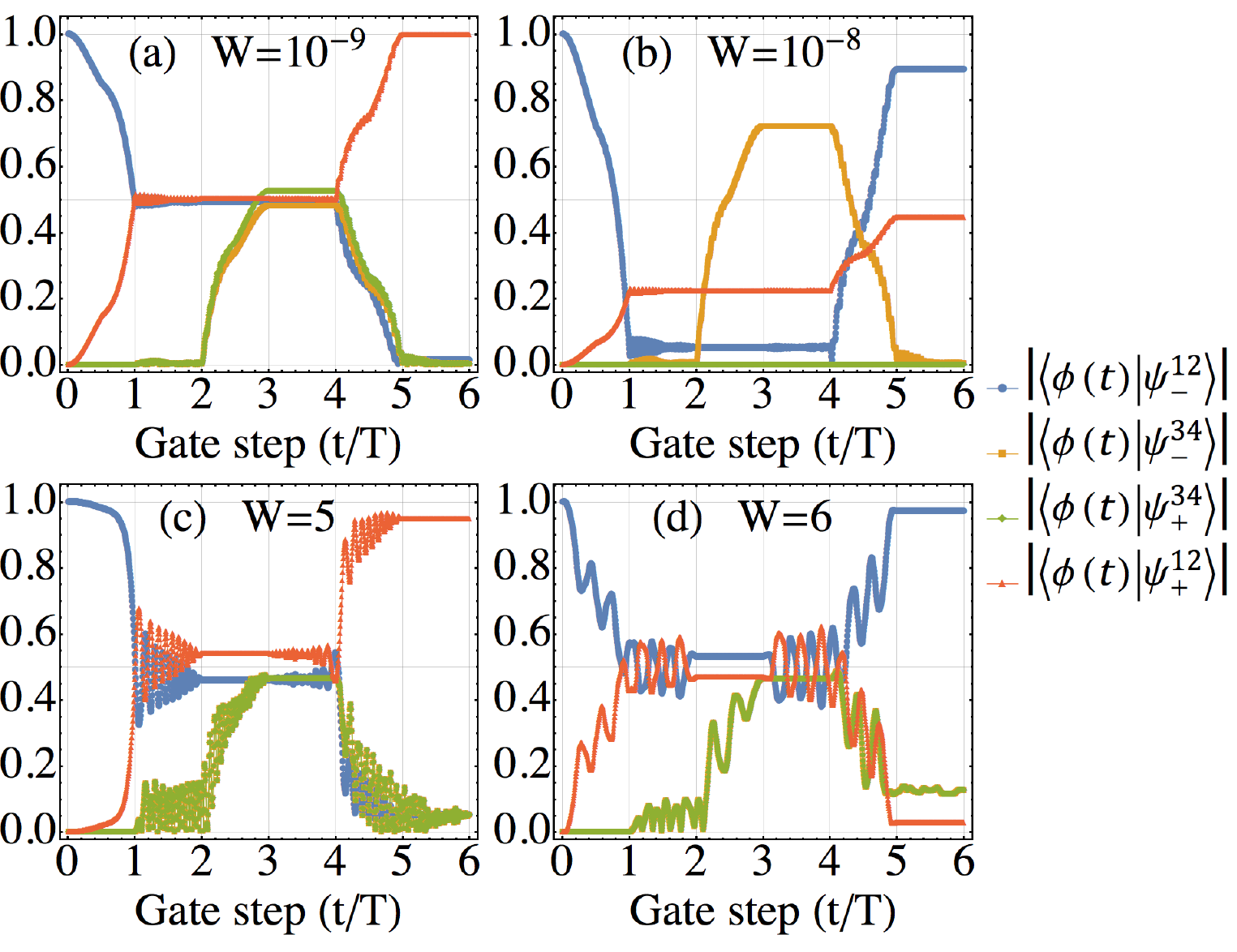}
\caption{Evolution of the eigenstate $| \phi(t) \rangle$ as two Majorana zero-modes $\gamma_2$ and $\gamma_3$ are swapped twice in the presence of disorder. Each gate step takes time of $T=100$, SC gap $\Delta_{\mathrm{SC}} \approx 2.6$, and the coupling energy between Majorana zero-modes $\epsilon_{12}, \epsilon_{34} \approx 1 \times 10^{-9}$. (a), (b) PH symmetry breaking disorder $H_{\mathrm{dis}}$ with disorder strength (a) $W=10^{-9}$, and (b) $W=10^{-8}$; (c), (d) PH symmetry conserved disorder $H_{\mathrm{dis}}^{\mathcal{P}}$ with disorder strength (c) $W=5$, and (d) $W=6$. The non-Abelian braiding properties that $\psi_{-}^{12} \to \psi_{+}^{12}$ is nearly well-preserved in (a) and (c), while destructed in (b) and (d).}
\label{disorder_p-wave_SC}
\end{figure}

Similar to the Jackiw-Rebbi case, there are six Majorana zero-modes (denoted as $\gamma_{i=1,2,...,6}$) in the cross-shaped junction, and the effective Hamiltonian describing the coupling energy ($\epsilon_{2i-1,2i}$) and the ``fictitious'' energy deviation ($\Delta_{2i-1,2i}$) of the Majorana zero-modes reads \cite{Chui-Zhen_cross_junction_supp}:

\begin{equation}
\label{Majorana_effective_Hamiltonian}
H_{\mathrm{M}} = i\epsilon_{12} \gamma_1 \gamma_2 + i\epsilon_{34} \gamma_4 \gamma_3 
+ \Delta_{12} \gamma_1 \gamma_1 - \Delta_{12} \gamma_2 \gamma_2 + \Delta_{34} \gamma_3 \gamma_3 - \Delta_{34} \gamma_4 \gamma_4
\end{equation}

\noindent where the two widely separated Majorana zero-modes $\gamma_5$ and $\gamma_6$ are neglected. Two eigenstates of Eq. (\ref{Majorana_effective_Hamiltonian}) formed by $\gamma_1$ and $\gamma_2$ are $\psi_{\pm}^{12} = \frac{1}{\sqrt{2}C_{12}^{\pm}} \{ \gamma_1 + i[{\widetilde\Delta}_{12} \mp (1+\widetilde{\Delta}_{12}^2)^{1/2} ]\gamma_2 \}$ (where $\widetilde{\Delta}_{12} \equiv \Delta_{12}/\epsilon_{12}$, and $C_{12}^{\pm}$ are normalization constants). Swapping two Majorana zero-modes $\gamma_2$ and $\gamma_3$ lead to $\gamma_2 \to \gamma_3$ and $\gamma_3 \to -\gamma_2$ \cite{IvanovPRL2001_supp}, therefore a full braiding process swapping $\gamma_2$ and $\gamma_3$ twice gives rise to $\gamma_2 \to -\gamma_2$ and $\gamma_3 \to -\gamma_3$. In the presence of the degeneracy between Majorana zero-modes ($\widetilde{\Delta}_{12}=0$), the eigenstates $\psi_{-}^{12} = \frac{1}{\sqrt{2}} (\gamma_1 + i\gamma_2) = (\psi_{+}^{12})^{\dagger}$, hence the braiding operation results in $\psi_{-}^{12} |G\rangle \to \psi_{+}^{12} |G\rangle = (\psi_{-}^{12})^{\dagger}|G\rangle$ as $|G\rangle$ indicates the SC ground state. On the contrary, assume that the degeneracy is lifted (i.e. $\widetilde{\Delta}_{12} \neq 0$), then the braiding properties $\gamma_{2}\to-\gamma_{2}$ lead to 

\begin{equation}
\label{fidelity_Majorana}
\left| \langle \phi(t=6T) | \psi_{+}^{12} \rangle \right| = \left( 1+\widetilde{\Delta}_{12}^{2} \right)^{-1/2}
\end{equation}

\noindent [where $| \phi(t=0) \rangle = | \psi_{-}^{12} \rangle$], whose form is exactly the same as the braiding ``fidelity'' of Jackiw-Rebbi zero-modes [Eq. (\ref{fidelity}) in the main text].

\begin{figure}[t]
    \centering
    \includegraphics[width=0.425\textwidth]{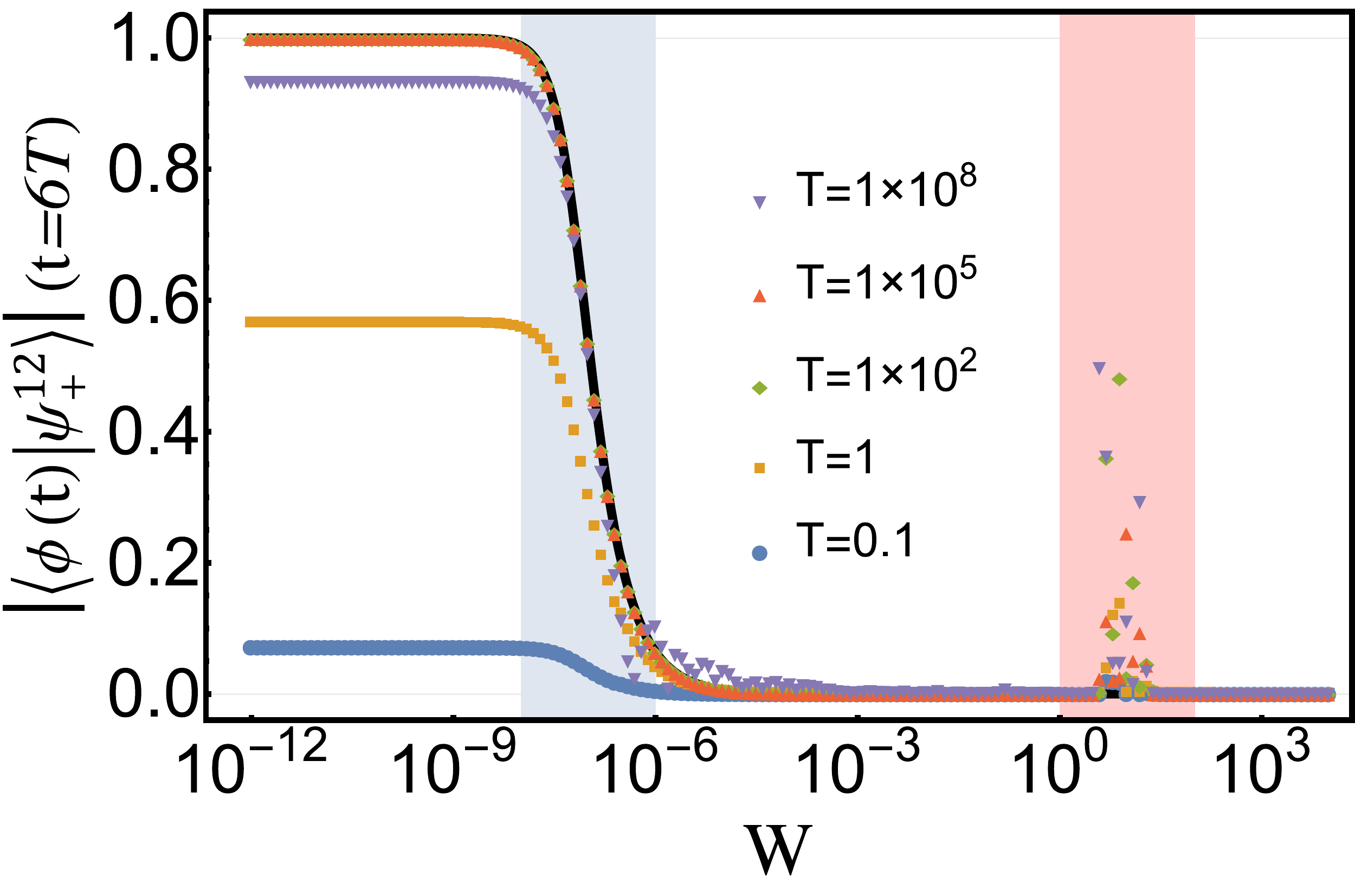}
    \caption{Braiding ``fidelity'' defined as Eq. (\ref{fidelity_Majorana}) in a fixed disorder profile with different disorder strength $W$ and braiding time $T$. The coupling energy between Majorana zero-modes $\epsilon_{12}, \epsilon_{34} \approx 1 \times 10^{-9}$, and the SC gap $\Delta_{\mathrm{SC}} \approx 2.6$. Too long ($T=1 \times 10^8$) or too short ($T=0.1$) braiding time will violate the adiabatic condition $ \Delta_{\mathrm{SC}} \gg 1/T \gg \epsilon_{12}, \epsilon_{34}$. The fitting curve of Eq. (\ref{fidelity_Majorana}) is shown in black. The blue (red) shaded region indicates $W \sim \epsilon_{12}, \epsilon_{34}$ ($W \sim \Delta_{\mathrm{SC}}$).}			
\label{fidelity_plot}
\end{figure}

The braiding of Majorana zero-modes in the $p \pm ip$-wave SC and the corresponding evolution of the eigenstates can be numerically simulated in the same way as the Jackiw-Rebbi case. Though it might be experimentally unrealistic, a ``fictitious'' PH symmetry breaking disorder in the form of $H_{\mathrm{dis}} = \mathrm{diag} \{ V_1(\mathbf{r}), V_2(\mathbf{r}) \}$ with $V_i(\mathbf{r}) \in [-W/2, W/2]$ still can be introduced in the numerical simulation, and the non-Abelian braiding properties are destructed by weak disorder $W \ll \Delta_{\mathrm{SC}}$ [Fig. \ref{disorder_p-wave_SC} (a), (b), where $\Delta_{\mathrm{SC}}$ is the SC gap]. Besides, numerical results of the braiding ``fidelity'' [Eq. (\ref{fidelity_Majorana})] in a fixed disorder profile with different disorder strength $W$ and braiding time $T$ is shown as Fig. \ref{fidelity_plot}. Too long (too short) braiding time $T$ will induce energy level transition between different Majorana zero-modes (between Majorana zero-modes and bulk states) as the energy level transition $(\epsilon_{\mathrm{f}} -\epsilon_{\mathrm{i}}) \sim 1/T $. For intermiediate braiding time satisfying the adiabatic condition $ \Delta_{\mathrm{SC}} \gg 1/T \gg \epsilon_{12}, \epsilon_{34}$, the braiding results can be perfectly fitted by Eq. ({\ref{fidelity_Majorana}}) as $ \left| \langle \phi(t=6T) | \psi_{+}^{12} \rangle \right| = \left( 1+ a W^{2} \right)^{-1/2} $ (black curve in Fig. \ref{fidelity_plot}, $a$ is a fitting constant) since $\widetilde{\Delta}_{12} \propto W$ in the fixed disorder profile.

Conversely, in the condition that the disorder has a PH symmetry conserving form as $H_{\mathrm{dis}}^{\mathcal{P}} = V_1(\mathbf{r})\sigma_z$ [satisfying $ -H_{\mathrm{dis}}^{\mathcal{P}} = \mathcal{P} (H_{\mathrm{dis}}^{\mathcal{P}})^T \mathcal{P}^{-1}$], Majorana zero-modes' non-Abelian properties are well preserved until the SC gap is destructed by strong disorder $W \sim \Delta_{\mathrm{SC}}$ [Fig. \ref{disorder_p-wave_SC} (c), (d)]. It is reasonable since $H_{\mathrm{dis}}^{\mathcal{P}}$ imposes disorder of opposite signs on the electron band and hole band of the $p \pm ip$-wave SC, therefore the energy deviation of Majorana zero-modes ($\Delta_{2i-1,2i}$) vanishes and the non-Abelian properties maintain integrity.




\end{document}